\documentclass[12pt]{article}

\usepackage{amsmath}
\usepackage{amsfonts}
\usepackage{amssymb}
\usepackage{graphics,graphicx,tikz}
\usepackage{subfig}
\usepackage{graphicx}
\usepackage{soul}
\usepackage{cite}
\usepackage{graphicx}
\usetikzlibrary{patterns}
\usepackage{enumitem}
\usepackage{array}
\usepackage{amsbsy}
\usepackage{bbold}
\usepackage{appendix}
\usepackage{mathrsfs}
\usepackage{physics}
\usepackage{amsthm}
\usepackage{cancel}
\usepackage{yfonts}
\usepackage{inputenc}
\usepackage{chngcntr}

\numberwithin{equation}{section} 
\usepackage[linktocpage]{hyperref}
\hypersetup{
	colorlinks=true,
	linkcolor=blue,
	filecolor=magenta,      
	urlcolor=blue,
	citecolor=blue
}
\def\beq{\begin{equation}}
\def\eeq{\end{equation}}
\newcommand{\commentOut}[1]{}
\def\bea{\begin{align}}
\def\eea{\end{align}}

\usepackage{color}

\begin{document}
\begin{titlepage}
\vskip 0.1cm
\vskip 1.5cm
\begin{flushright}
\end{flushright}
\vskip 1.0cm
\begin{center}
{\Large \bf Kerr binary dynamics from minimal coupling and double copy}
\vskip 1.0cm {\large  Francesco Alessio$^{a,b}$} \\[0.7cm]

{\it \small $^a$ NORDITA, KTH Royal Institute of Technology and Stockholm University, \\
 Hannes Alfv{\'{e}}ns v{\"{a}}g 12, SE-11419 Stockholm, Sweden  }\\
 {\it \small $^b$ Department of Physics and Astronomy, Uppsala University,\\ Box 516, SE-75120 Uppsala, Sweden}
\\

\end{center}
\begin{abstract}
We construct a new Yang-Mills Lagrangian based on a notion of minimal coupling that incorporates classical spin effects. The construction relies on the introduction of a new covariant derivative, which we name ``classical spin covariant derivative", that is compatible with the three-point interaction of the $\sqrt{\mathrm{Kerr}}$ solution with the gauge field. The resulting Lagrangian, besides the correct three-point coupling, predicts a unique choice for contact terms and therefore it can be used to compute higher-point amplitudes such as the Compton, unaffected by spurious poles. Using double copy techniques we use this theory to extract gravity amplitudes and observables that are relevant to describe Kerr binary dynamics to all orders in the spin. In particular, we compute the 2PM ($\mathcal{O}(G^2_N)$) $2\rightarrow 2$ elastic scattering amplitude between two classically spinning objects to all orders in the spin and use it to extract the 2PM scattering angle. 
\end{abstract}
\end{titlepage}


\section{Introduction}
\label{intro}
The idea that black holes share certain features with elementary particles is not new and, in fact, it has quite a long history (see \textit{e.g.}\cite{tHooft:1984kcu,Holzhey:1991bx,Monteiro:2014cda}). One of the reasons that makes these two classes of apparently unrelated objects similar is that they are both characterized by few charges, namely the mass, the intrinsic angular momentum \textit{i.e.} the spin and the electric charge. Although being exact solutions of the full, non-linear Einstein equations, black holes seem to be simple objects, at least if compared to other solutions such as neutron stars.

Beside being interesting in itself, the aforementioned analogy can be concretely used to perform analytical precision computations relevant in the context of gravitational waves physics, that has recently dragged a lot of attention because of the many detections made by LIGO, Virgo and Kagra of waves emitted by strongly gravitating binary systems \cite{LIGOScientific:2016aoc,LIGOScientific:2016sjg,LIGOScientific:2017bnn,LIGOScientific:2017ycc,LIGOScientific:2017vwq,LIGOScientific:2018mvr,LIGOScientific:2020ibl}
. During the very first phase of the coalescence of two black holes, known as the inspiral phase, the two compact objects are widely separated and the gravitational interaction between them is weak. In this regime one can neglect the internal structure of the bodies, sized by the Schwarzschild radius, and successfully use perturbation theory to extract the perturbative expansion in the gravitational coupling constant of the observables one is interested in. This is called post-Minkowskian (PM) regime and it is usually approached either by using classical general relativity (GR) methods or scattering amplitude techniques in quantum field theory (QFT)
\cite{Buonanno:1998gg,Goldberger:2004jt,Goldberger:2016iau,Cachazo:2017jef,Luna:2017dtq,Bjerrum-Bohr:2018xdl,Cheung:2018wkq,Kosower:2018adc,Bern:2019nnu,Bern:2019crd,Brandhuber:2019qpg,AccettulliHuber:2019jqo,KoemansCollado:2019ggb,Cristofoli:2019neg,Bjerrum-Bohr:2019kec,Cheung:2020gyp,Parra-Martinez:2020dzs,Brandhuber:2021eyq,Bern:2021dqo,Cristofoli:2020uzm,AccettulliHuber:2020dal,delaCruz:2020bbn,Cristofoli:2021vyo,Herrmann:2021tct,Cristofoli:2021jas,Damour:2017zjx,Herrmann:2021lqe,DiVecchia:2020ymx,DiVecchia:2021ndb,DiVecchia:2021bdo,DiVecchia:2022nna,DiVecchia:2022piu,Bjerrum-Bohr:2021vuf,Bjerrum-Bohr:2021din,Bjerrum-Bohr:2021wwt,Bern:2021yeh,Kalin:2020mvi,Kalin:2020fhe,Dlapa:2021npj,Dlapa:2021vgp,Kalin:2022hph,Dlapa:2022lmu,Mogull:2020sak,Jakobsen:2022psy,Khalil:2022ylj,Jones:2022aji,Bini:2022wrq,Brandhuber:2023hhy,Georgoudis:2023lgf}. However, because of on-shell techniques and many simplifications, state-of-the-art theoretical predictions are made within the latter approach, where one treats the two gravitating bodies as elementary point-like particles, computes a scattering amplitude and then extracts its classical limit, which roughly amounts to sending $\hbar$ to zero, after having correctly restored it in the theory. Eventually open orbits observables, obtained by differentiating the classical scattering amplitude, can be mapped into bounded orbits ones \cite{Kalin:2019rwq,Kalin:2019inp,Cho:2021arx}
 and can be compared to those derived in the more familiar Post-Newtonian (PN) regime (see \textit{e.g.} \cite{Blanchet:2006zz} for a review), where, together with the weak field approximation, one also considers small velocities. 

It is not surprising that the theory used to describe the gravitational interaction between Schwarzschild black holes, only characterized by their masses, is in the PM regime simply that of minimally coupled massive scalars to gravity\footnote{In this framework, adding non-minimal terms to the Lagrangian means considering tidal deformations contributions.}. Computations can be significantly simplified using double copy \cite{Kawai:1985xq,Bern:2008qj,Bern:2010ue,Bern:2019prr,Plefka:2019wyg} and, infact, it has been shown that scalar QCD provides the single copy theory to compute the relevant building blocks and amplitudes for the the more complicated theory of gravity \cite{Johansson:2014zca,Johansson:2015oia,
Luna:2017dtq,Monteiro:2014cda,Luna:2016due,Lee:2018gxc,Monteiro:2020plf,Monteiro:2021ztt}. State-of-the-art results for the scattering between two spinless black holes in both the conservative and radiative sectors are up to 4PM ($\mathcal{O}(G^4_N)$). 

However, more realistic black holes are also characterized by the spin vector $S^{\mu}$ \cite{Kerr:1963ud}. Because of this additional scale the problem is much more complicated and, on top of each PM order, there is a spin multipole expansion in the classical spin vector that, in general, depends on the nature of the compact body one is describing \cite{Arkani-Hamed:2017jhn,Guevara:2017csg,Bini:2017xzy,Vines:2017hyw,Bini:2018ywr,Vines:2018gqi,Guevara:2018wpp,Chung:2018kqs,Bautista:2019tdr,Bautista:2019evw,Maybee:2019jus,Guevara:2019fsj,Arkani-Hamed:2019ymq,Johansson:2019dnu,Chung:2019duq,Damgaard:2019lfh,Chung:2019yfs,Aoude:2020onz,Chung:2020rrz,Bern:2020buy,Aoude:2020ygw,Guevara:2020xjx,Liu:2021zxr,Kosmopoulos:2021zoq,Aoude:2021oqj,Jakobsen:2021lvp,Bautista:2021wfy,Chiodaroli:2021eug,Haddad:2021znf,Gonzo:2021drq,Jakobsen:2021zvh,Saketh:2021sri,Adamo:2021rfq,Chen:2021qkk,Jakobsen:2022fcj,Aoude:2022trd,Aoude:2022thd,Bern:2022kto,Alessio:2022kwv,Chen:2022clh,Ochirov:2022nqz,Damgaard:2022jem,FebresCordero:2022jts,Menezes:2022tcs,Riva:2022fru,Cangemi:2022abk,Hung:2022azf,Jakobsen:2022zsx,Saketh:2022wap,Bjerrum-Bohr:2023jau,Bautista:2022wjf,Cangemi:2022bew,Comberiati:2022ldk,Comberiati:2022cpm,Kim:2023drc,A:2022wsk,Hoogeveen:2023bqa,Haddad:2023ylx,Elkhidir:2023dco}. Quantum-mechanically, the spin vector satisfies $S^{\mu}S_{\mu}=-\hbar^2s(s+1)$, where $s$ is the spin quantum number. Hence, when trying to attack the classical scattering problem using QFT methods, it seems natural to make use of spinning massive fields, with higher and higher value of the spin quantum number $s$ , in such a way to keep the value of $S^{\mu}S_{\mu}$ finite when sending $\hbar$ to zero. This will yield a spin multipole expansion in the form of a polynomial truncated at order $(S^{\mu})^{2s}$. The idea of using quantum spinning fields to mimic Kerr black holes is corroborated by the ``minimally coupled"\footnote{Here the notion of ``minimal coupling" does not have anything to do with the QFT one, which we will exploit later on in the paper.} three-point amplitudes presented in \cite{Arkani-Hamed:2017jhn} that, in the classical limit, match the classical computation of the linearised stress-energy tensor of a Kerr black hole \cite{Vines:2017hyw}. The latter resums all the classical spin dependence of the amplitude into a simple and compact exponentiated form that can be interpreted, on the support of the three-point kinematics, as an exponentiated version of the subleading soft graviton theorem \cite{Guevara:2018wpp,Cachazo:2014fwa}. Using this three-point amplitude, the 1PM classical amplitude for the scattering between two Kerr black holes has been easily constructed to all order in spin \cite{Guevara:2018wpp,Chung:2018kqs,Guevara:2019fsj}. However, going to 2PM requires the knowledge of the tree-level Compton amplitude and, when one constructs such object using BCFW recursion relations \cite{Britto:2005fq}, the opposite-helicity sector is affected, for $s>2$, \textit{i.e.} for higher spins, by the presence of unphysical spurious poles. The same problem appears for gauge theories when $s>1$, when starting from the $\sqrt{\mathrm{Kerr}}$ three-point amplitude.
There has been a lot of effort to understand how to remove these poles. The Lagrangian approaches of \cite{Chiodaroli:2021eug,Bern:2022kto} give healthy Compton amplitudes up to $s=5/2$.
Assuming that the Compton amplitude additionally satisfies a ``shift-symmetry" to all orders in spin leads to a result unaffected by the above-mentioned spurious poles \cite{Aoude:2022trd}. A microscopical reason for assuming such symmetry is, however, far from being clear. All these efforts and proposals should eventually be compared with solutions of the classical Teukolsky equation, recently found in \cite{Bautista:2022wjf}.

In this paper, we follow a different path. We try to answer the question of whether it is possible, in principle, to construct an effective Lagrangian field theory describing the interactions of Kerr black holes with gravity depending on the \textit{classical spin} $S^{\mu}$ and therefore carrying, as a built-in feature, an infinite spin multipole expansion. There are several motivations to ask this question. The first is of purely theoretical nature. As argued above, black holes are, from a certain perspective, simple and this simplicity, for the Schwarzschild case, is reflected in the standard QFT notion of minimal coupling. On the other hand, Kerr black-holes look much more sophisticated but they are actually not, as clear in Kerr-Schild coordinates \cite{Kerr2}. Indeed, Kerr spacetime can be entirely reconstructed by performing a Newman-Janis transformation \cite{Newman:1965tw} to the radial coordinate characterizing the spinless solution\footnote{A discussion on Kerr-Schild coordinates can be found in appendix \ref{S2}.}. Constructing a new field theory for Kerr by implementing this transformation directly at the level of the Lagrangian would represent at least a good candidate to describe Kerr black holes as elementary particles. A second, but not less important, motivation concerns the frontier of precision computations. Having a Lagrangian built upon the criterion of simplicity would make it possible to proceed beyond the first PM order in the perturbative expansion of gravitational observables. One could argue that, unless there is a classical GR computation that counterparts the analysis of this paper, these observables could have little to do with Kerr black holes. However, in any case, the theory we present here could always be regarded as a ``minimal" model on top of which one is free to add operators of higher order in the curvature, opportunely multiplied by Wilson coefficients, in order to better and better approximate true classically spinning solutions. 

In practice we do not construct here the theory of gravity, but we limit ourselves to build its single copy, gauge theory counterpart. Then, implementing a form of the double copy that adapts to classically spinning theories, we obtain gravity amplitudes. 

The paper is organized as follows. We start in section \ref{S3} where, after discussing the usual notion of minimal coupling and its role in scalar QED, we construct a new theory by introducing a covariant derivative which is compatible with the $\sqrt{\mathrm{Kerr}}$ three-point amplitude and that carries information about its interaction with the gauge field. We show explicit expressions for the Lagrangian and for the contact terms, automatically generated by this  definition of minimal coupling. The Newman-Janis shift, discussed in some detail in appendix \ref{S2}, is a constitutive element of our Lagrangian. In section \ref{S4} we consider tree-level amplitudes and we show that, on-shell, spinning amplitudes involving an arbitrary number of massless legs entirely factorize in terms of spinless ones and that the full spin contribution can remarkably be encoded into a simple exponential factor. Because they come from a local Lagrangian, such amplitudes will be unaffected by unphysical poles. The difference between the opposite-helicity Compton amplitude we find within our approach and the ``minimally coupled" one is spelled out in appendix \ref{AC1}. Using double copy for classically spinning theories, in section \ref{S5}, we obtain gravity amplitudes relevant to describe Kerr black holes binary dynamics. Section \ref{S6} is devoted to the computation of the 2PM amplitude, eikonal and scattering angle using on-shell techniques. Integrals appearing in the one-loop amplitude and eikonal are computed in appendix \ref{1-loop} and \ref{AC}. We conclude in \ref{S7} with a discussion. Throughout the paper we use mostly minus signature.

\section{Minimal coupling and classical spin}
\label{S3}
In this section we start by considering scalar QED, review its Feynman rules and compute the well-known Compton amplitude. Afterwards, by analogy and requiring compatibility with the three-point coupling of the $\sqrt{\mathrm{Kerr}}$ solution, we build a new Lagrangian describing classically spinning scalars coupled to the electromagnetic field.  
\subsubsection*{Scalar QED}
\label{SS31}
A scalar $\phi$ with mass $m$ and electric charge $e$ minimally coupled to the electromagnetic field is described by the scalar QED Lagrangian,
\begin{align}
\label{F13}
\mathcal{L}=-\frac{1}{4}F_{\mu\nu}F^{\mu\nu}+(D_{\mu}\phi)^*(D^{\mu}\phi)-m^2|\phi|^2,\hspace{0.5cm} D_{\mu}=\partial_{\mu}+ieA_{\mu}.
\end{align}
Under gauge transformations,
\begin{align}
\label{F14}
A_{\mu}\rightarrow A_{\mu}-\frac{i}{e}U\partial_{\mu}U^{*},\hspace{1cm}\phi\rightarrow U\phi,\hspace{1cm}U=e^{ie\hspace{0.05cm}\chi}\in U(1),
\end{align}
the covariant derivative transforms as $D_{\mu}\phi=UD_{\mu}\phi$. This is the usual QFT notion of minimal coupling, that consists in providing interactions between matter and gauge fields with the least number of derivatives in a gauge invariant way. Clearly other non-minimal interaction terms, containing more derivatives, may be added to the Lagrangian, but we will not consider them here. Beside ensuring gauge invariance, the replacement $\partial_{\mu}\phi\rightarrow D_{\mu}\phi$ in the free Lagrangian automatically yields three- and four-point couplings,
\begin{align}
\label{F15}
\mathcal{L}_{3 \mathrm{pt}}=ieA^{\mu}(\phi\partial_{\mu}\phi^*-\phi^*\partial_{\mu}\phi),\qquad \mathcal{L}_{4 \mathrm{pt}}=e^2A^{\mu}A_{\mu}|\phi|^2,
\end{align}
from which we easily read the Feynman rules,
\begin{equation}
\label{F16a}
\vcenter{\hbox{\includegraphics[width=3.6cm,height=4cm]{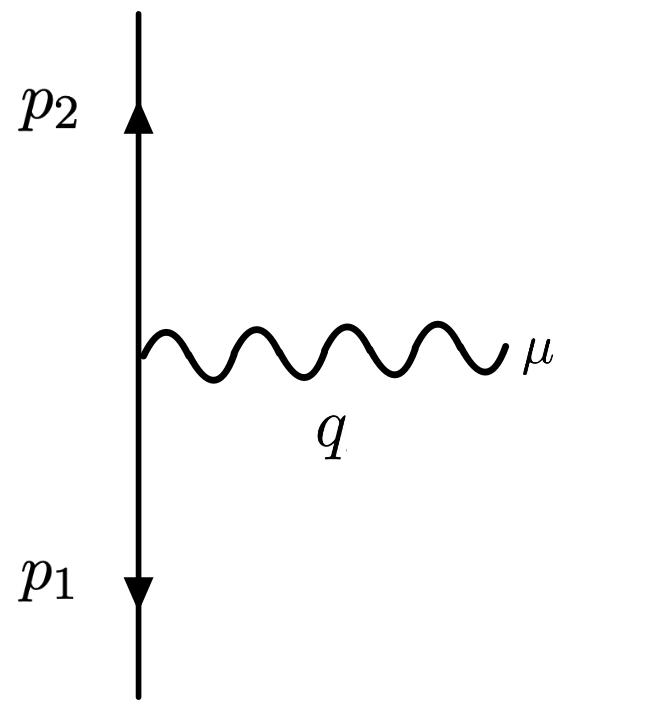}}}
\begin{aligned}
\hspace{0.1cm}=-ie\hspace{0.05cm}p^{\mu},\end{aligned}
\begin{aligned}\hspace{2cm}
\vcenter{\hbox{\includegraphics[width=3.6cm,height=4cm]{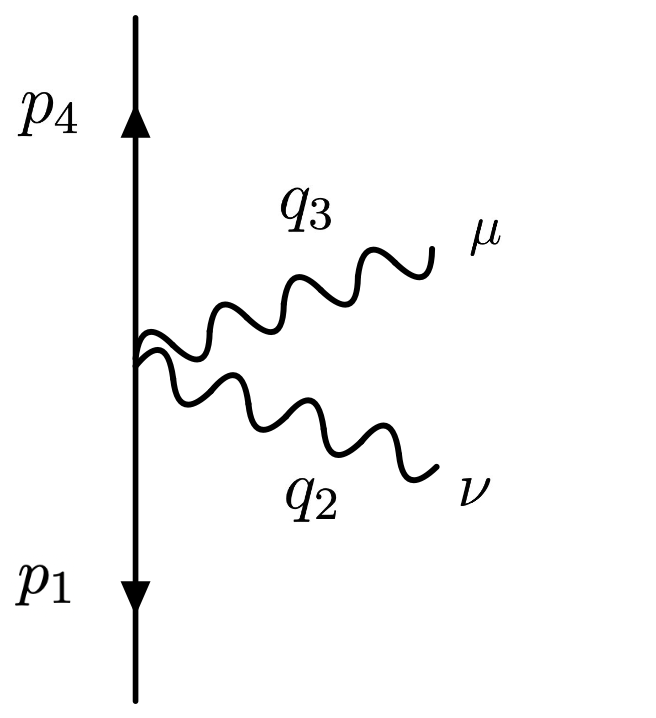}}}
\hspace{0.1cm}=2ie^2 \eta^{\mu\nu},
\end{aligned}
\end{equation}
where $\phi$ ($\phi^*$) has momentum $p_1$ ($p_2$), $p \equiv p_2-p_1$ and where, by total momentum conservation for the first diagram $p_1+p_2+q=0$ and for the second $p_1+q_1+q_2+p_4=0$\footnote{In our convention all momenta are outgoing.}. The tree-level Compton amplitude is the sum of the three diagrams below,
\begin{equation}
\label{F16}
i\mathcal{A}^{\mu\nu}_4(p_i,q_i)=\vcenter{\hbox{\includegraphics[width=2.9cm,height=3.9cm]{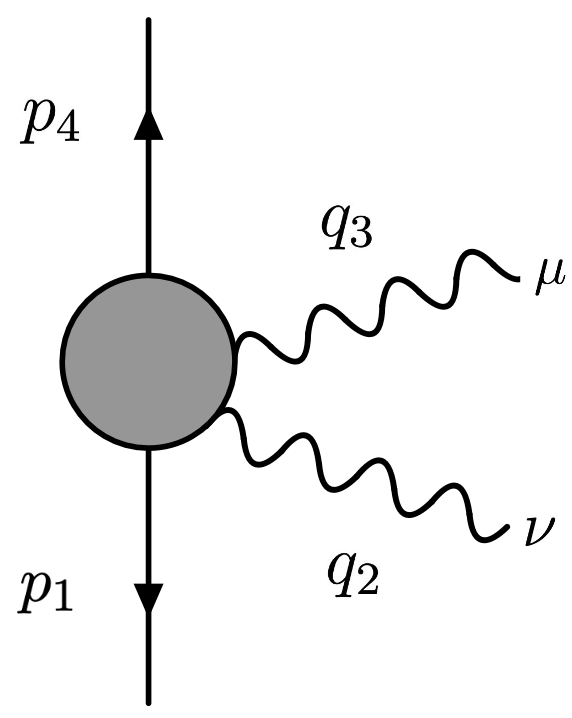}}}
\begin{aligned}
\hspace{0.1cm}=\end{aligned}
\vcenter{\hbox{\includegraphics[width=2.5cm,height=3.9cm]{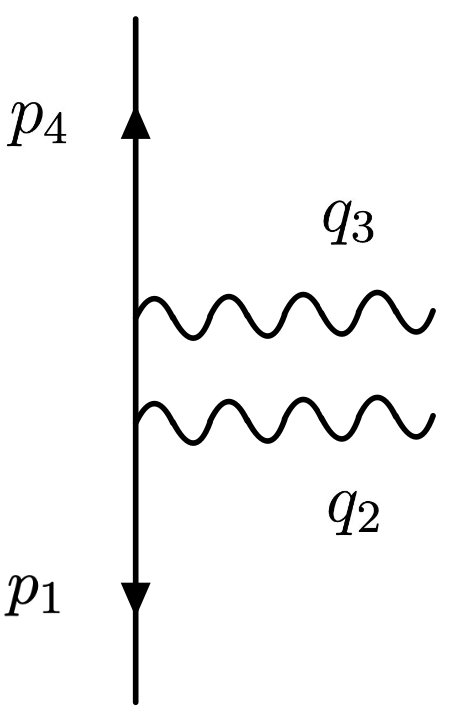}}}
\hspace{0.1cm}\begin{aligned}+\end{aligned}\vcenter{\hbox{\includegraphics[width=2.5cm,height=3.9cm]{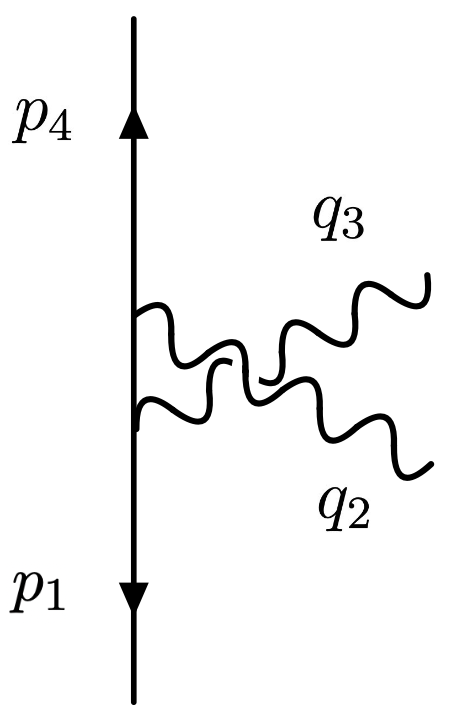}}}
\hspace{0.1cm}\begin{aligned}+\end{aligned}\vcenter{\hbox{\includegraphics[width=2.5cm,height=3.9cm]{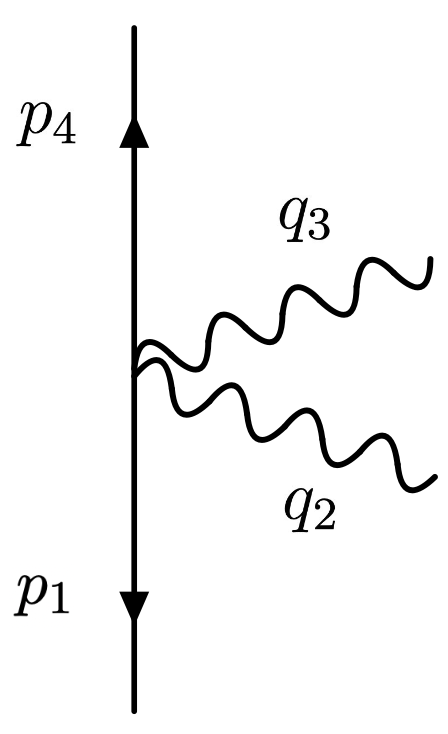}}}.
\end{equation}
Using the scalar propagator, we get for $\mathcal{A}_4^{\mu\nu}$,
\begin{align}
\label{F17}
\mathcal{A}^{\mu\nu}_4(p_i,q_i)=-2e^2\bigg(\frac{p_1^{\mu}p_1^{\nu}+q_2^{\mu}p_1^{\nu}}{p_1\cdot q_2}+\frac{p_1^{\mu}p_1^{\nu}+p_1^{\mu}q_3^{\nu}}{p_1\cdot q_3}-\eta^{\mu\nu}\bigg).
\end{align}
Contracting with physical polarization vectors satisfying $\varepsilon_i^2=0=\varepsilon_i\cdot q_i$ and using the on-shell conditions $q^2_i=0$ we get,
\begin{align}
\label{F18}
\mathcal{A}_4(p_i,q_i)\equiv \varepsilon_{1\nu}\varepsilon_{2\mu}\mathcal{A}^{\mu\nu}_4(p_i,q_i)=-\frac{2e^2}{p_1\cdot q_2 p_1\cdot q_3}p_1\cdot F_2\cdot F_3\cdot p_1,
\end{align}
where we introduced the field strength $F_i^{\mu\nu}=2\varepsilon_i^{[\mu}q_i^{\nu]}$\footnote{Symmetrization and antisymmetrization are defined as $A^{(\mu}B^{\nu)}=\frac{1}{2}(A^{\mu}B^{\nu}+A^{\nu}B^{\mu})$ and $A^{[\mu}B^{\nu]}=\frac{1}{2}(A^{\mu}B^{\nu}-A^{\nu}B^{\mu})$.} and used that, on-shell, $p_1\cdot(q_2+q_3)=-q_2\cdot q_3$. Note that $\mathcal{A}_4$ is gauge invariant because it is written purely in terms of gauge invariant field strengths. If we knew only the three-point coupling in equation \eqref{F15} \textit{i.e.} if we summed only the first two diagrams in \eqref{F16}, we could have guessed the four-point contact term, necessary to achieve gauge invariance, just by requiring that $\delta_{\chi}\mathcal{A}_4=0$. This suggests that, in a minimally coupled gauge theory, contact terms are always generated as a consequence of the three-point coupling that, in turn, comes from the notion of a ``good" covariant derivative.

As widely discussed in the literature and in the introduction, amplitudes describing Schwarzschild black holes come from minimally coupled scalars to gravity. Such minimal coupling is, at the gauge theory level, encoded in the notion of covariant derivative and indeed, the classical Compton amplitude for Schwarzschild black holes can be simply obtained by double copying $\mathcal{A}_4$ in \eqref{F18}. This shows that the simplicity of black holes is intimately connected to the notion of minimal coupling. In the following we will \textit{assume}, relying on the considerations of the previous section and in particular on the Newman-Janis algorithm, discussed in appendix \ref{S2}, that Kerr black holes should also be described by a minimally coupled, simple, theory. In other words, we will ask whether the classical three-point amplitude for $\sqrt{\mathrm{Kerr}}$, known to all orders in the spin, yields a good covariant derivative and can therefore be used to generate contact terms in the same way they are derived in the spinless case. The answer to this question, quite surprisingly, is yes. 
\subsubsection*{$\sqrt{\mathrm{Kerr}}$ and classically spinning gauge theory}
\label{SS32}
In the worldline picture, the Coulomb field can be thought of as generated by a charged scalar sourced by a static point particle and hence it is a solution of the Maxwell equations,
\begin{align}
\label{F19}
\partial_{\mu}F^{\mu\nu}(x)=J^{\nu}(x),\qquad J^{\mu}(x)=e \int \frac{d\tau}{2\pi} u^{\mu}\delta^{(4)}(x-u \tau),
\end{align}
where $\tau$ is the proper time of the point particle moving on the worldline $x(\tau)$ and $u^{\mu}(\tau)=\dot{x}(\tau)$ is the unit tangent vector to the curve satisfying $u_{\mu}u^{\mu}=1$ (see \textit{e.g.} \cite{Luna:2016due}). Taking the Fourier transform of $J^{\mu}(x)$ yields,
\begin{align}
\label{F20}
\tilde{J}^{\mu}(q)=\int\frac{d^4 x}{(2\pi)^4}e^{iq\cdot x}J^{\mu}(x)=e\hspace{0.05cm}u^{\mu}\delta(q\cdot u).
\end{align}
From a Lagrangian field theory point of view, the coupling of the charged particle to the electromagnetic field is encoded in the term $A_{\mu}J^{\mu}$ and therefore the Feynman rule for the three-point amplitude is the Fourier transform $\tilde{J}^{\nu}(q)$ of $J^{\nu}(x)$. Indeed, introducing $p^{\mu}=mu^{\mu}$ the Fourier transform \eqref{F20} is proportional to the first equation in \eqref{F16a}. The additional delta function appearing in \eqref{F20} enforces three-point kinematics $p_1\cdot q=p_2\cdot q=p\cdot q=0$.

As shown in the context of gravity in \cite{Vines:2017hyw}, the $\sqrt{\mathrm{Kerr}}$ solution, which is the electromagnetic field generated by a rotating disc of radius $a$, is sourced by the following current,
\begin{align}
\label{F21}
J^{\mu}(x)=e\int\frac{d\tau}{2\pi}\mathrm{exp}\{\epsilon^{\mu}{}_{\nu}(a,\partial)\}u^{\nu}\delta^{(4)}(x-u\tau),\qquad \epsilon^{\mu}{}_{\nu}(a,\partial)\equiv\epsilon^{\mu}{}_{\nu\rho\sigma}a^{\rho}\partial^{\sigma},
\end{align}
where the rescaled spin vector $a^{\mu}$ is related to the classical Pauli-Lubanski spin vector $S^{\mu}$ by $S^{\mu}=m a^{\mu}$ and satisfies the Tulczjew-Dixon SSC condition $a\cdot u=0$ \cite{SSC1,SSC2}. This time, performing the Fourier transform yields \cite{Guevara:2018wpp}, 
\begin{align}
\label{F22}
\tilde{J}^{\mu}(q)=\int\frac{d^4 x}{(2\pi)^4}e^{iq\cdot x}J^{\mu}(x)=e\hspace{0.05cm}\mathrm{exp}\{i\epsilon^{\mu}{}_{\nu}(a,q)\}u^{\nu}\delta(q\cdot u), 
\end{align}
which leads to the following three-point amplitude,
\begin{equation}
\label{F23}
\vcenter{\hbox{\includegraphics[width=3.6cm,height=4cm]{3pt.png}}}
\begin{aligned}
\hspace{0.1cm}=-ie\hspace{0.05cm}\mathrm{exp}\{i\epsilon^{\mu}{}_{\nu}(a,q)\}p^{\nu}\equiv -ie\hspace{0.05cm}\Lambda^{\mu}{}_{\nu}(a,q)p^{\nu}.\end{aligned}
\end{equation}
In practice, switching from scalar QED to $\sqrt{\mathrm{Kerr}}$ amounts to multiplying the three-point Feynman rule of the former by a  matrix $\Lambda^{\mu}{}_{\nu}(a,q)$, which is the exponential of the antisymmetric tensor $\omega_{\mu\nu}=-\omega_{\nu\mu}=\epsilon_{\mu\nu}(a,q)$. Remarkably, it was shown in \cite{Guevara:2018wpp,Guevara:2019fsj} that the amplitude \eqref{F23} corresponds to the classical limit of ``minimally coupled" three-point amplitudes constructed in \cite{Arkani-Hamed:2017jhn}. Indeed, contracting \eqref{F23} with $\varepsilon_{\mu}$ and expanding the exponential gives,
\begin{align}
\label{F23a}
\nonumber &-ie\hspace{0.05cm}\varepsilon^{\mu}\bigg[\eta_{\mu\nu}+i\epsilon_{\mu\nu}(a,q)-\frac{1}{2!}\epsilon_{\mu\gamma}(a,q)\epsilon^{\gamma}{}_{\nu}(a,q)+\dots\bigg]p^{\nu}\\=&ie\hspace{0.05cm}\varepsilon\cdot p\bigg[1+i\frac{\varepsilon\cdot S\cdot q}{\varepsilon\cdot p}+\frac{1}{2}(a\cdot q)^2+\dots \bigg]=
ie\hspace{0.05cm}\varepsilon\cdot p\hspace{0.05cm}\mathrm{exp}\left\{\frac{i}{2}\frac{S^{\mu\nu}F_{\mu\nu}}{\varepsilon\cdot p}\right\},
\end{align}
where $S^{\mu\nu}=\epsilon^{\mu\nu\alpha\beta}p_{\alpha}a_{\beta}$ is the spin tensor of the massive particle and where we repeatedly used the relations,
\begin{align}
\label{F23ab}
\epsilon^{\mu\nu\rho\sigma}\epsilon_{\alpha\beta\gamma\delta}=-4!\delta^{[\mu}_{\alpha}\delta^{\nu}_{\beta}\delta^{\rho}_{\gamma}\delta^{\sigma]}_{\delta},\qquad\epsilon^{\mu\nu\rho\lambda}\epsilon_{\alpha\beta\gamma\lambda}=-3!\delta^{[\mu}_{\alpha}\delta^{\nu}_{\beta}\delta^{\rho]}_{\gamma},
\end{align}
together with, crucially, the three-point kinematics $p\cdot q=0$, the on-shell conditions $\varepsilon^2=q^2=\varepsilon\cdot q=0$ and the SSC $a\cdot p=0$ as well. It is suggestive to note \cite{Guevara:2018wpp} that the exponential factor appearing in \eqref{F23a} is exactly an exponentiated version of the subleading soft photon theorem \cite{Low:1958sn}. The poles appearing when expanding the exponential in \eqref{F23a} for $a>1$ are cancelled due to three-point kinematics,
\begin{align}
\label{F23b}
\frac{i}{2}\frac{S^{\mu\nu}F_{\mu\nu}}{\varepsilon\cdot p}=i\frac{\varepsilon\cdot S\cdot q}{\varepsilon\cdot p}=-\frac{a\cdot\varepsilon\hspace{0.05cm}p\cdot q-a\cdot q\hspace{0.05cm}p\cdot\varepsilon}{p\cdot\varepsilon}=a\cdot q,
\end{align}
where we used $\varepsilon\cdot S\cdot q=\sqrt{(\varepsilon\cdot S\cdot q)^2}$ and the Gram determinants in \eqref{F23ab} with the assumption of real polarization vector $\varepsilon$. 

Asking that the structure appearing in \eqref{F23} comes from a standard QFT notion of minimal coupling, \textit{i.e.} from the existence of a covariant derivative, implies that the latter should be,
\begin{align}
\label{F24}
D_{\mu}^{(a)}=\partial_{\mu}-ie\hspace{0.05cm}\mathrm{exp}\{\epsilon^{\nu}{}_{\mu}(a,\partial)\}A_{\nu}.
\end{align}
Under gauge transformations as in \eqref{F14} we have,
\begin{align}
\label{F25}
D_{\mu}^{(a)}\phi\rightarrow &\partial_{\mu}(U\phi)-ie\hspace{0.05cm}\mathrm{exp}\{\epsilon^{\nu}{}_{\mu}(a,\partial)\}(A_{\nu}+\partial_{\nu}\chi)(U\phi)=UD_{\mu}^{(a)}\phi,
\end{align}
where we used that the matrix exponential acts trivially on pure gauge configurations, $\mathrm{exp}\{\epsilon^{\nu}{}_{\mu}(a,\partial)\}\partial_{\nu}\chi=\partial_{\mu}\chi$.  Equation \eqref{F25} shows that the operator $D_{\mu}^{(a)}$ is indeed a good covariant derivative and it can be used to construct consistent interactions between matter and gauge fields. We will denote $D_{\mu}^{(a)}$ as c\textit{lassical spin covariant derivative}. We are naturally led to consider the gauge invariant Lagrangian,
\begin{align}
\label{F26}
\mathcal{L}=-\frac{1}{4}F_{\mu\nu}F^{\mu\nu}+(D^{(a)}_{\mu}\phi)^*(D^{\mu}_{(a)}\phi)-m^2|\phi|^2=\mathcal{L}_{\mathrm{free}}+\mathcal{L}_{3\mathrm{pt}}+\mathcal{L}_{4\mathrm{pt}},
\end{align}
where,
\begin{align}
\label{F27}
&\mathcal{L}_{3 \mathrm{pt}}=ie\hspace{0.05cm}\mathrm{exp}\{\epsilon_{\mu\nu}(a,\partial)\}A^{\mu}(\phi\partial^{\nu}\phi^*-\phi^*\partial^{\nu}\phi),\\\nonumber\\&\label{F28}\mathcal{L}_{4 \mathrm{pt}}=e^2\mathrm{exp}\{\epsilon_{\mu\rho}(a,\partial)\}A^{\mu}\mathrm{exp}\{\epsilon_{\nu}{}^{\rho}(a,\partial)\}A^{\nu}|\phi|^2,
\end{align}
that automatically yields the three-point diagram in \eqref{F23} and the four-point contact term,
\begin{equation}
\label{F28}
\begin{aligned}
\vcenter{\hbox{\includegraphics[width=3.6cm,height=4cm]{ct.png}}}
\hspace{0.1cm}=2ie^2 \Lambda^{\mu}{}_{\rho}(a,q_3)\Lambda^{\nu\rho}(a,q_2).
\end{aligned}
\end{equation}
Before proceeding further we note here that, in general, the field strength is defined as the commutator of covariant derivatives, $F_{\mu\nu}=\frac{i}{e}[D_{\mu},D_{\nu}]$. Using this definition, together with the classical spin connection in \eqref{F24}, would change the photon propagator. However, in the theory \eqref{F26} we are keeping the free part as it is in the standard YM Lagrangian, without modifying the field strength as $F_{\mu\nu}\rightarrow F^{(a)}_{\mu\nu}$. In other words, the photon field gets dressed with the spin exponential matrix \textit{only} when it interacts \textit{locally} with the massive, classically spinning field $\phi$. Away from it photons are not affected by the spin. This remark is not important in the abelian version of the theory we are presenting here, because none of the diagrams appearing in the Compton \eqref{F16} contains a photon propagator. When extending this theory to gluons, however, beside the two massive channels and the contact term, there is also a massless channel involving the gluon propagator. 

It comes as a consequence of the above considerations that the spinning Compton amplitude $\mathcal{A}_4^{\mu\nu}(a)$ is simply factorized in terms of the spinless one as,
\begin{align}
\label{F29}
\mathcal{A}_4^{\mu\nu}(a;p_i,q_i)=\Lambda^{\mu}{}_{\rho}(a,q_3)\Lambda^{\nu}{}_{\sigma}(a,q_2)\mathcal{A}_4^{\rho\sigma}(p_i,q_i).
\end{align}
We need to contract with polarization vectors $\varepsilon_{2\nu}$ and $\varepsilon_{3\mu}$ and hence it is convenient to define new vectors $\varepsilon'^{\mu}_{i}=\Lambda_{\nu}{}^{\mu}(a,q_i)\varepsilon^{\nu}_i$ so that,
\begin{align}
\label{F30}
\mathcal{A}_4(a;p_i,q_i)=\varepsilon_{2\nu}\varepsilon_{3\mu}\mathcal{A}_4^{\mu\nu}(a;p_i,q_i)=-\frac{2e^2}{p_1\cdot q_2 p_1\cdot q_3}p_1\cdot F'_2\cdot F'_3\cdot p_1,
\end{align}
where $F_{i}'^{\mu\nu}=2\varepsilon'^{[\mu}_iq_i^{\nu]}$. In general, all the spinning tree-level amplitudes $\mathcal{A}_{N}^{\mu_2\dots\mu_{N-1}}(a)$ involving two massive and $N-2$ massless lines with momenta $q_2,\dots,q_{N-1}$ following from the Lagrangian \eqref{F26} can be written in terms of the spinless ones as,
\begin{equation}
\label{F31}
\mathcal{A}^{\mu_2\dots\mu_{N-1}}_N(a;p_i,q_i)=\vcenter{\hbox{\includegraphics[width=3.4cm,height=4cm]{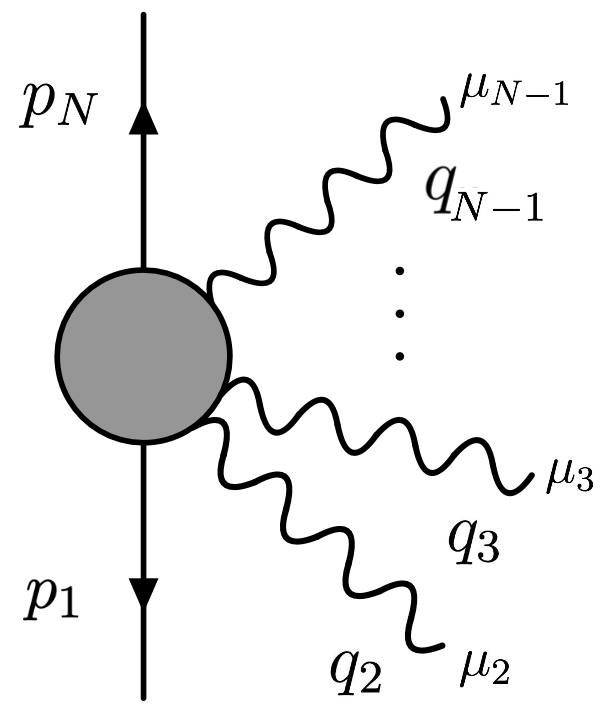}}}\\
\begin{aligned}
\hspace{0.1cm}= \Lambda^{\mu_2}{}_{\nu_2}(a,q_2)\dots&\Lambda^{\mu_{N-1}}{}_{\nu_{N-1}}(a,q_{N-1})\\&\mathcal{A}_{N}^{\nu_2\dots\nu_{N-1}}(p_i,q_i).\end{aligned}
\end{equation}
\section{Spin exponentiation}
\label{S4}
The Lagrangian presented in the previous section predicts tree-level spinning amplitudes containing various factors of $\Lambda_{\nu}{}^{\mu}(a,q_i)\varepsilon^{\nu}_i$ for each external massless line. Here we would like to analyse these factors more in detail and to extract an alternative expression that possibly does not contain the Levi-Civita symbol. We introduce the shorthand notation $\epsilon^{\mu_1\dots\mu_n}{}_{\nu_1\dots\nu_n}v^{\nu_1}_1\dots v^{\nu_m}_m\equiv \epsilon^{\mu_1\dots \mu_n}(v_1,\dots,v_m)$ for $m$ vectors $v_1,\dots, v_m$, that generalizes the one used in \eqref{F21} and we start by expanding the exponential as,
\begin{align}
\label{F32}
\nonumber \mathrm{exp}\{i\epsilon_{\nu}{}^{\mu}(a,q)\}\varepsilon^{\nu}=\hspace{0.05cm}&\varepsilon^{\mu}-i\epsilon^{\mu}(\varepsilon,a,q)+\frac{1}{2!}\epsilon^{\alpha}(\varepsilon,a,q)\epsilon_{\alpha}{}^{\mu}(a,q)\\+&\frac{i}{3!}\epsilon^{\alpha}(\varepsilon,a,q)\epsilon_{\alpha}{}^{\beta}(a,q)\epsilon_{\beta}{}^{\mu}(a,q)+\dots
\end{align}
Defining the dual field strength $F^{*\mu\nu}_i=\frac{i}{2}\epsilon^{\mu\nu}{}_{\rho\sigma}F^{\rho\sigma}=i\epsilon^{\mu\nu}(\varepsilon,q)$ \footnote{Usually the dual field strength $F^{*}_{\mu\nu}$ is defined without the $i$ factor.} and using the on-shell equality,
\begin{align}
\label{F33}
\epsilon^{\alpha}(\varepsilon,a,q)\epsilon_{\alpha}{}^{\beta}(a,q)=a\cdot q(a\cdot q\varepsilon^{\beta}-a\cdot\varepsilon q^{\beta})=a\cdot q F^{\beta\rho}a_{\rho},
\end{align}
the series in \eqref{F32} can easily be resummed,
\begin{align}
\label{F34}
\nonumber \mathrm{exp}\{i\epsilon_{\nu}{}^{\mu}(a,q)\}\varepsilon^{\nu}=\hspace{0.05cm}&\varepsilon^{\mu}+F^{*\mu\rho}a_{\rho}+\frac{1}{2!}(a\cdot q_i)a_{\rho}F^{\mu\rho}+\frac{1}{3!}(a\cdot q)^2F^{*\mu\rho}a_{\rho}+\dots\\\nonumber\\=\hspace{0.05cm}&\frac{a\cdot \varepsilon}{a\cdot q}q^{\mu}+\frac{F^{*\mu\rho}a_{\rho}}{a\cdot q}\sinh(a\cdot q)+\frac{F^{\mu\rho}a_{\rho}}{a\cdot q}\cosh(a\cdot q).
\end{align}
So far, we have not specified the helicity of the massless state $\varepsilon$. A field strength admits a decomposition into its positive and negative helicity components $F^{\mu\nu}_{h}$ with $h=\pm$ \cite{Monteiro:2021ztt},
\begin{align}
\label{F35}
F^{+}_{\mu\nu}=\frac{1}{2}(F_{\mu\nu}-F^{*}_{\mu\nu}),\qquad F^{-}_{\mu\nu}=\frac{1}{2}(F_{\mu\nu}+F^{*}_{\mu\nu}),
\end{align}
that, plugged into \eqref{F34} yields,
\begin{align}
\label{F36}
\mathrm{exp}\{i\epsilon_{\nu}{}^{\mu}(a,q)\}\varepsilon^{\nu}=\frac{a\cdot \varepsilon}{a\cdot q}q^{\mu}+\frac{F_{+}^{\mu\rho}a_{\rho}}{a\cdot q}e^{-a\cdot q}+\frac{F_{-}^{\mu\rho}a_{\rho}}{a\cdot q}e^{a\cdot q}.
\end{align}
Therefore for fixed helicity $h$ we find,
\begin{align}
\label{F37}
\mathrm{exp}\{i\epsilon_{\nu}{}^{\mu}(a,q)\}\varepsilon^{\nu}_{h}=e^{-h a\cdot q}\varepsilon^{\mu}_{h}-\frac{(e^{-h a\cdot q}-1)}{a\cdot q}a\cdot \varepsilon_{h}q^{\mu},\qquad h=\pm.
\end{align}
We conclude that, up to a gauge transformation, positive and negative helicity vectors are eigenvectors of the operator $\mathrm{exp}\{i\epsilon_{\nu}{}^{\mu}(a,q)\}$ with eigenvalues $e^{-h a\cdot q}$.
An immediate consequence is that the spinning three-point amplitude in \eqref{F23}, on-shell ($p\cdot q=0$) and once specified the helicity of the emitted photon, reads simply $(p\cdot\varepsilon_{h})e^{-h a\cdot q}$. More in general, any tree-level gauge invariant amplitude of the form \eqref{F31} will contain a product of spin exponentials,
\begin{align}
\label{F38}
\nonumber \mathcal{A}^{h_2\dots h_{N-1}}_N(a;p_i,q_i)&=\epsilon^{h_2}_{\mu_2} \Lambda^{\mu_2}{}_{\nu_2}(a,q_2)\dots \epsilon^{h_{N-1}}_{\mu_{N-1}}\Lambda^{\mu_{N-1}}{}_{\nu_{N-1}}(a,q_{N-1})\mathcal{A}_{N}^{\nu_2\dots\nu_{N-1}}(p_i,q_i)\\\nonumber\\&=e^{-\sum_{i=2}^{N-1}h_i\hspace{0.01cm}a\cdot q_i}\mathcal{A}^{h_2\dots h_{N-1}}_N(p_i,q_i).
\end{align}
In particular, for $N=4$, \textit{i.e.} for the Compton amplitude, the equal- and opposite-helicity sectors are,
\begin{align}
\label{F39}
\mathcal{A}_4^{\pm\pm}(a;p_i,q_i)=e^{\mp a\cdot(q_2+q_3)}\mathcal{A}_4^{\pm\pm}(p_i,q_i),\quad\mathcal{A}_4^{\pm\mp}(a;p_i,q_i)=e^{\mp a\cdot(q_2-q_3)}\mathcal{A}_4^{\pm\mp}(p_i,q_i),
\end{align}
respectively. Notice that the opposite-helicity Compton amplitude in \eqref{F39} is ``shift symmetric" \cite{Aoude:2022trd}. Indeed, if we perform $\delta_{\xi}a=\xi (q_2+q_3)$ we have $\delta_{\xi}\mathcal{A}_4^{\pm\mp}(a)=\mathcal{A}_4^{\pm\mp}$ using $q_i^2=0$.

So far, we were only concerned about gauge theories. In section \ref{S5} we will introduce a notion of double copy for the theory just presented and we will use it to extract gravity amplitudes that will be the main building blocks of the 2PM computation we will present in section \ref{S6}.

However, before turning to gravity, we end this section with some considerations on the amplitudes in \eqref{F39} and their relation with the ones present in the literature. Using the kinematics specified in \eqref{F40}-\eqref{F43.1} we find that in the classical limit the Compton amplitudes in \eqref{F39} are,
\begin{align}
\label{F43}
\mathcal{A}_{4}^{\pm\pm}(a;p_i,q_i)=2e^2\sin^2(\theta/2)e^{\pm\mathcal{G}(\omega,a,\theta)},\quad \mathcal{A}_{4}^{\pm\mp}(a;p_i,q_i)=2e^2\cos^2(\theta/2)e^{\pm\mathcal{F}(\omega,a,\theta)},
\end{align}
with,
\begin{align}
\label{F44}
&\mathcal{G}(\omega,a,\theta)=\omega a^x\sin\theta-2\omega a^z\sin^2(\theta/2),\\ &\mathcal{F}(\omega,a,\theta)=-\omega a^x\sin\theta-2\omega a^z\cos^2(\theta/2).
\end{align}
Remarkably, the equal-helicity sector we find within our approach coincides with the one found in \cite{Saketh:2022wap,Bautista:2022wjf}, once we send $\omega\rightarrow-\omega$. Equivalently, it is compatible with the classical limit of the ``minimally coupled" amplitudes presented in \cite{Arkani-Hamed:2017jhn} in massive spinor helicity variables, obtained using BCFW recursion relations and sewing together three-point amplitudes in \eqref{F23} and \eqref{F23ab}. 
However, the spin exponential structure in the second of \eqref{F43} differs from the one in \cite{Arkani-Hamed:2017jhn,Bautista:2019tdr,Saketh:2022wap,Bautista:2022wjf,Kim:2023drc}, for $s<2$. In particular, our opposite-helicity Compton amplitude does not depend on the complex vector $w^{\mu}$ present  \textit{e.g.} in \cite{Aoude:2020onz,Saketh:2022wap}. Here we briefly analise the mismatch and we show that it is related to the lack of factorization of our Compton amplitude on the massless channel, which is absent in the abelian version of the theory presented here. Indeed, it easy to see that, while the equal-helicity sector correctly factorizes as the product of a three-point amplitude in \eqref{F23} and a three-gluon amplitude, the opposite-helicity one does not. 

At four points there are three independent momenta and therefore the complex  vector $w^{\mu}$ can be decomposed as, 
\begin{align}
\label{A5}
w^{\mu}=\alpha_1q_2^{\mu}+\alpha_2q_3^{\mu}+\alpha_3 p_1^{\mu}+\alpha_4\epsilon^{\mu}(q_2,q_3,p_1).
\end{align}
Imposing $w\cdot q_2=w\cdot q_3=w_{\mu}w^{\mu}=0$ fixes $\alpha_1,\alpha_2$ and $\alpha_4$ in terms of $\alpha_3$ as
\begin{align}
\label{A6}
\alpha_1=-\frac{p_1\cdot q_3}{q_2\cdot q_3}\alpha_3,\qquad \alpha_2=-\frac{p_1\cdot q_2}{q_2\cdot q_3}\alpha_3, \qquad \alpha_4=\pm \frac{i}{q_2\cdot q_3}\alpha_3.
\end{align}
Therefore, we define (note that $a\cdot p_1=0$ because of SSC so we do not display the term proportional to $p_1^{\mu}$),
 \begin{align}
\label{A7}
w^{\mu}_{+-}=-\frac{\alpha_3}{q_1\cdot q_2}[p_1\cdot q_2q_1^{\mu}+p_1\cdot q_1 q_2^{\mu}+i\epsilon^{\mu}(q_1,q_2,p_1)],\qquad w^{\mu}_{-+}=(w^{\mu}_{+-})^*.
\end{align}
We then fix the coefficient $\alpha_3$ by comparison with equation (3.1) of \cite{Aoude:2022trd} and we find,
\begin{align}
\label{A16}
\alpha_3=\frac{q_1\cdot q_2 p_1\cdot (q_2-q_1)}{(2p_1\cdot q_2 p_1\cdot q_1-q_1\cdot q_2 m^2)}.
\end{align}
We can then proceed construct the amplitudes,
\begin{align}
\label{A17}
\tilde{\mathcal{A}}^{+-}_4(a;p_i,q_i)=e^{-a\cdot w_{+-}}\mathcal{A}^{+-}_4(a;p_i,q_i),\qquad\tilde{\mathcal{A}}^{+-}_4(a;p_i,q_i)=e^{a\cdot w_{-+}}\mathcal{A}^{-+}_4(a;p_i,q_i)
\end{align}
The feature of these amplitudes is that now they present correct factorization properties on the massless channel. Indeed as $t=2q_1\cdot q_2\rightarrow 0$,
\begin{align}
\label{A18}
\tilde{\mathcal{A}}^{\pm\mp}_4(a;p_i,q_i)|_{t\rightarrow 0}=e^{- a\cdot (q_1+q_2)}\mathcal{A}^{\pm\mp}_4(p_i,q_i)|_{t\rightarrow 0}. 
\end{align}
as they should. Using the kinematics specified in \eqref{F40}-\eqref{F43.1} we find that in the classical limit the amplitudes in \eqref{A17} can be written as,
\begin{align}
\label{A19}
\tilde{\mathcal{A}}_4^{+-}(a;p_i,q_i)=2e^2\cos^2(\theta/2)e^{\mathcal{\tilde{F}}_{+-}(\omega,a,\theta)},\qquad\tilde{\mathcal{A}}_4^{-+}(a;p_i,q_i)=2e^2\cos^2(\theta/2)e^{\mathcal{\tilde{F}}_{-+}(\omega,a,\theta)},
\end{align}
where,
\begin{align}
\label{A20}
&\tilde{\mathcal{F}}_{+-}(\omega,a,\theta)=\omega(a^x+a^z\sin\theta+2ia^y-a^x\cos\theta)\tan(\theta/2),\\
&\tilde{\mathcal{F}}_{-+}(\omega,a,\theta)=-\omega(a^x+a^z\sin\theta-2ia^y-a^x\cos\theta)\tan(\theta/2)=\tilde{\mathcal{F}}^*_{+-}(-\omega,a,\theta),
\end{align}
and they match, for instance, the ones in \cite{Bautista:2022wjf}.

\section{Double copy for classical spin}
\label{S5}
In this section we will use the gauge theory just discussed to get gravity amplitudes. In order to do so, we will implement the double copy for classically spinning particles,
\begin{align}
\label{F53}
\mathrm{Kerr}(a)=\sqrt{\mathrm{Kerr}(a)}\otimes\sqrt{\mathrm{Kerr}(0)}.
\end{align}
The first simple example where it is possible to verify that this version of the double copy is satisfied is at three-point. The classical gravity amplitude involving two Kerr black holes and a graviton was found in \cite{Vines:2017hyw} and it reads,
\begin{align}
\label{F54}
\mathcal{M}_3^{\mu\nu}(a;p,q)=i\kappa\hspace{0.01cm} p^{(\mu}\mathrm{exp}\{i\epsilon^{\nu)}{}_{\rho}(a,q)\}p^{\rho},\qquad \kappa=\sqrt{32\pi G_N},
\end{align}
and it can easily be obtained using the prescription in \eqref{F53},
\begin{align}
\label{F55}
\mathcal{M}_3^{\mu\nu}(a;p,q)=-i\mathcal{A}_3^{(\mu}(p,q)\mathcal{A}^{\nu)}_3(a;p,q)|_{e\rightarrow \sqrt{\kappa}}.
\end{align}
Let us now turn to the 1PM elastic scattering amplitude between two Kerr black holes with rescaled spin vectors $a_1$ and $a_2$. We can consider the product,
\begin{align}
\label{F56}
\nonumber q^2\mathcal{A}^{1\mathrm{PM}}(a;p_i,q)\mathcal{A}^{1\mathrm{PM}}(p_i,q)=&\frac{(e_1m_1)^2(e_2m_2)^2\sigma^2}{4 q^2}\bigg[\sum_{\pm}(1\pm v)^2e^{\pm i\frac{\epsilon(p_1,p_2,q,a)}{m_1m_2\sigma v}}\\&+2(1-v^2)\cos(\frac{\epsilon(p_1,p_2,q,a)}{m_1m_2\sigma v})\bigg],
\end{align}
where $\mathcal{A}^{1\mathrm{PM}}(a)$ is the classical 1PM amplitude for the elastic $2\rightarrow 2$ scattering. It has been calculated in appendix \ref{AB} and it is explicitly given in \eqref{FA6}. In \eqref{F56} $a=a_1+a_2$, $\sigma=p_1\cdot p_2/m_1m_2$ and $v=\sqrt{\sigma^2-1}/\sigma$. We see that the first term in the above square bracket will reproduce, once reinserting all numerical factors, the classical amplitude for the one graviton exchange of two Kerr black holes shown in \cite{Guevara:2019fsj}. However there is another term, whose origin can be traced back to the fact that when we naively double copy gauge theory amplitudes having internal massless propagators, the result, together with the graviton contribution, will contain also that of the dilaton and the axion \footnote{The axion will only appear when the matter content of the theory has non-vanishing spin.}. Indeed, writing explicitly the product in \eqref{F56} yields,
\begin{align}
\label{F57}
\nonumber q^2&\mathcal{A}^{1\mathrm{PM}}(a;p_i,q)\mathcal{A}^{1\mathrm{PM}}(p_i,q)=\frac{1}{q^2}\mathcal{A}^{\mu}_3(a_1;p_1,q)\mathcal{A}^{\nu}_3(a_2;p_2,-q)\mathcal{A}^{\rho}_3(p_1,q)\mathcal{A}_3^{\sigma}(p_2,-q)\eta_{\mu\nu}\eta_{\rho\sigma}\\&=\frac{1}{q^2}\mathcal{A}^{\mu}_3(a_1;p_1,q)\mathcal{A}^{\nu}_3(a_2;p_2,-q)\mathcal{A}^{\rho}_3(p_1,q)\mathcal{A}_3^{\sigma}(p_2,-q)(\Pi^{(h)}_{\mu\rho;\nu\sigma}+\Pi^{(b)}_{\mu\rho;\nu\sigma}+\Pi^{(\phi)}_{\mu\rho;\nu\sigma}),
\end{align}
where we defined,
\begin{align}
\label{F58}
&\Pi^{(h)}_{\mu\rho;\nu\sigma}=\frac{1}{2}(\eta_{\mu\nu}\eta_{\rho\sigma}+\eta_{\mu\sigma}\eta_{\rho\nu}-\eta_{\mu\rho}\eta_{\nu\sigma}),\\
&\Pi^{(b)}_{\mu\rho;\nu\sigma}=\frac{1}{2}(\eta_{\mu\nu}\eta_{\rho\sigma}-\eta_{\mu\sigma}\eta_{\rho\nu}),\\
&\Pi^{(\phi)}_{\mu\rho;\nu\sigma}=\frac{1}{2}\eta_{\mu\rho}\eta_{\nu\sigma},
\end{align}
the projectors onto graviton, axion and dilaton, respectively. Therefore, in order to get the pure gravity result, we have to subtract the last two terms in \eqref{F57} to \eqref{F56}. By direct computation, the axion contribution is,
\begin{align}
\label{F59}
\nonumber \frac{1}{q^2}\mathcal{A}^{\mu}_3(a_1;p_1,q)&\mathcal{A}^{\nu}_3(a_2;p_2,-q)\mathcal{A}^{\rho}_3(p_1,q)\mathcal{A}_3^{\sigma}(p_2,-q)\Pi^{(b)}_{\mu\rho;\nu\sigma}\\&=-\frac{(e_1m_1)^2(e_2 m_2)^2}{2q^2}\sin(\frac{\epsilon(p_1,p_2,q,a_1)}{m_1m_2\sigma v})\sin(\frac{\epsilon(p_1,p_2,q,a_2)}{m_1m_2\sigma v}),
\end{align}
whereas that of the dilaton is,
\begin{align}
\label{F60}
\nonumber \frac{1}{q^2}\mathcal{A}^{\mu}_3(a_1;p_1,q)&\mathcal{A}^{\nu}_3(a_2;p_2,-q)\mathcal{A}^{\rho}_3(p_1,q)\mathcal{A}_3^{\sigma}(p_2,-q)\Pi^{(\phi)}_{\mu\rho;\nu\sigma}\\&=\frac{(e_1m_1)^2(e_2 m_2)^2}{2q^2}\cos(\frac{\epsilon(p_1,p_2,q,a_1)}{m_1m_2\sigma v})\cos(\frac{\epsilon(p_1,p_2,q,a_2)}{m_1m_2\sigma v}).
\end{align}
Notice that the axion contribution vanishes as the spin goes to zero. Subtracting \eqref{F59} and \eqref{F60} to \eqref{F56} yields the full 1PM gravity amplitude,
\begin{align}
\label{F61}
\mathcal{M}^{\mathrm{1PM}}(a;p_i,q)=\frac{\kappa^2}{4}\frac{m_1^2m_2^2\sigma^2}{ q^2}\sum_{\pm}(1\pm v)^2e^{\pm i\frac{\epsilon(p_1,p_2,q,a)}{m_1m_2\sigma v}},
\end{align}
where we replaced $e_i\rightarrow\sqrt{\kappa}$. Notice that the previous formula admits an interpretation in terms of the Newman-Janis shift discussed in detail in appendix \ref{S2}. Indeed, in order to extract observables out of $\mathcal{M}^{1\mathrm{PM}}(a)$, we need to go to impact parameter space and compute the eikonal, given by the Fourier transform,
\begin{align}
\label{F61.a}
\nonumber\tilde{\mathcal{M}}^{1\mathrm{PM}}(\mathbf{a};p_i,\mathbf{b})&=\int\frac{d^{4-2\epsilon}q}{(2\pi)^{4-2\epsilon}}\mathcal{M}^{1\mathrm{PM}}(a;p_i,q)\hat{\delta}(2p_1\cdot q)\hat{\delta}(2p_2\cdot q)e^{iq\cdot  b},\\&=\frac{1}{4m_1m_2\sqrt{\sigma^2-1}}\int\frac{d^{2-2\epsilon}\mathbf{q}}{(2\pi)^{2-2\epsilon}}\mathcal{M}^{1\mathrm{PM}}(\mathbf{a};p_i,\mathbf{q})e^{-i\mathbf{q}\cdot \mathbf{b}},
\end{align}
where we introduced the infrared regulator $\epsilon=(4-D)/2$, $\hat{\delta}(x)=2\pi\delta(x)$ and where $\mathbf{q}$ and $\mathbf{a}$ are the spatial component of the transfer momentum $q$ and of the rescaled spin vector $a$. Plugging \eqref{F61} into \eqref{F61.a} gives,
\begin{align}
\label{F62.a}
\nonumber\tilde{\mathcal{M}}^{1\mathrm{PM}}(\mathbf{a};p_i,\mathbf{b})&=\frac{\kappa^2m_1m_2\sigma^2}{64\sqrt{\sigma^2-1}}\frac{\Gamma(-\epsilon)}{\pi^{1-\epsilon}}\sum_{\pm}\frac{(1\pm v)^2}{\abs{\mathbf{b}\pm i\mathbf{a}}^{-2\epsilon}}\\&=\frac{\kappa^2m_1m_2\sigma^2}{64\sqrt{\sigma^2-1}}\frac{\Gamma(-\epsilon)}{\pi^{1-\epsilon}}\bigg[\frac{(1+v)^2}{\abs{\mathbf{b}}^{-2\epsilon}}+\frac{(1-v)^2}{\abs{\mathbf{b}^*}^{-2\epsilon}}\bigg]\bigg|_{\mathbf{b}\rightarrow\mathbf{b}+i\mathbf{a}},
\end{align}
which is reminiscent of a Newman-Janis shift (see \textit{e.g.} \eqref{F7}) of the spinless result. Therefore, the effect of the classical spin covariant derivative in \eqref{F23} on 1PM amplitudes in impact parameter space and hence on spinning observables, is equivalent to an implementation of such shift at the level of the Lagrangian.

Similarly, we can now consider the gravity Compton amplitude. In general, there is also a massless channel contributing to the latter and hence we should double copy a non abelian Yang-Mills Compton instead of \eqref{F39}. However, in the classical limit, gravitons self-interactions are subleading \cite{Monteiro:2020plf} and therefore we can simply use spinning QED Compton amplitudes constructed previously. At four-points the double copy reads,
\begin{align}
\label{F62}
\nonumber\mathcal{M}_4^{h_2,h_3}(a;p_i,q_i)&=\frac{\kappa^2}{e^4}\frac{8p_1\cdot q_2p_1\cdot q_3}{q_2\cdot q_3}\mathcal{A}_4^{h_2,h_3}(a;p_i,q_i)\mathcal{A}_4^{h_2,h_3}(p_i,q_i)\\&=\frac{\kappa^2}{2}\frac{1}{q_2\cdot q_3p_1\cdot q_2p_1\cdot q_3}e^{-a\cdot(h_2 q_2+h_3 q_3)}(p_1\cdot F_2^{h_2}\cdot F_3^{h_3}\cdot p_1)^2.
\end{align}
Again, the equal-helicity sector is reproduced exactly to all orders in the spin, differently from the opposite-helicity one. Using the complex vectors $w_{\pm\mp}$  in \eqref{A7}, up to $a^4$, the Kerr gravity Compton amplitude are given by,
\begin{align}
\label{F63}
\tilde{\mathcal{M}}_4^{+-}(a;p_i,q_i)=e^{-a\cdot w_{\pm}}\mathcal{M}_4^{+-}(a;p_i,q_i),\quad \tilde{\mathcal{M}}_4^{-+}(a;p_i,q_i)=e^{a\cdot w_{\mp}}\mathcal{M}_4^{-+}(a;p_i,q_i).
\end{align}
From $a^5$ on these amplitude are affected by spurious poles and therefore they are not physical.
\section{Classical 1-loop amplitude and 2PM observables}
\label{S6}
\begin{figure}[h]
        \centering
       {
            \includegraphics[width=.21\linewidth]{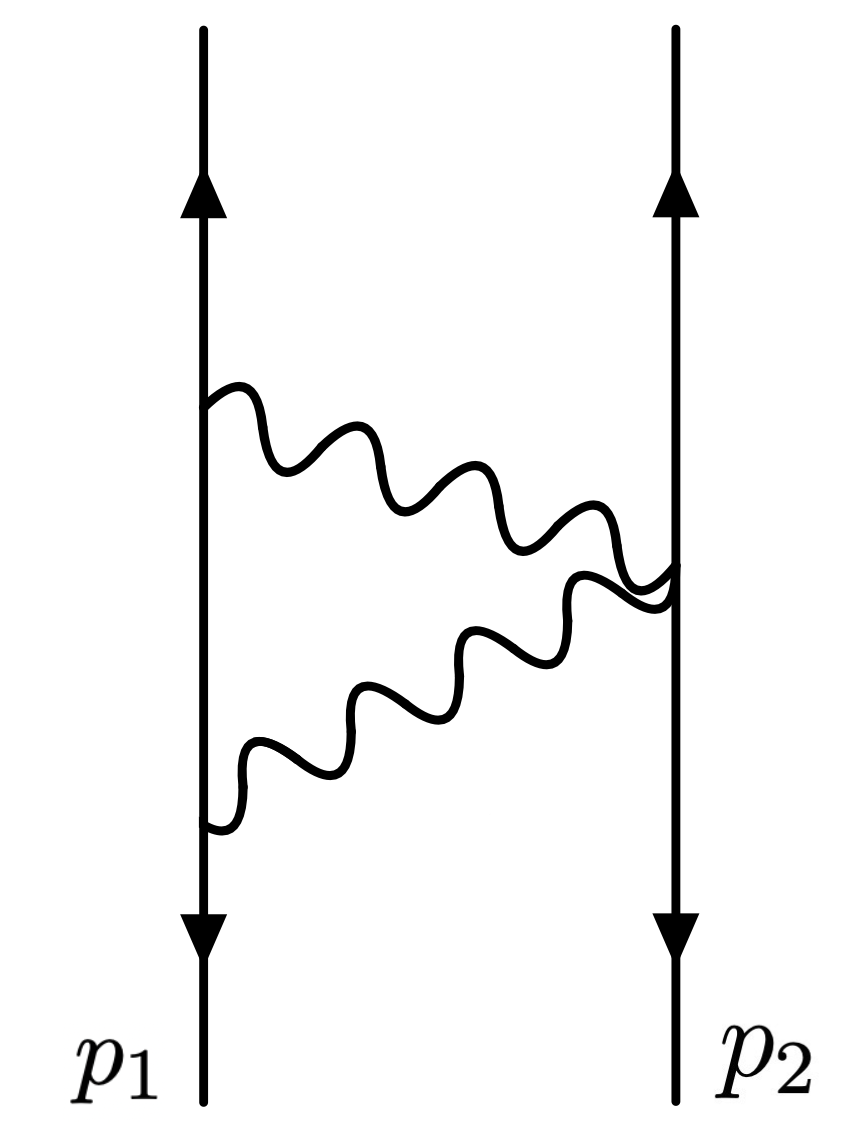}
            \label{subfig:A}
        }\qquad\qquad
        {
            \includegraphics[width=.21\linewidth]{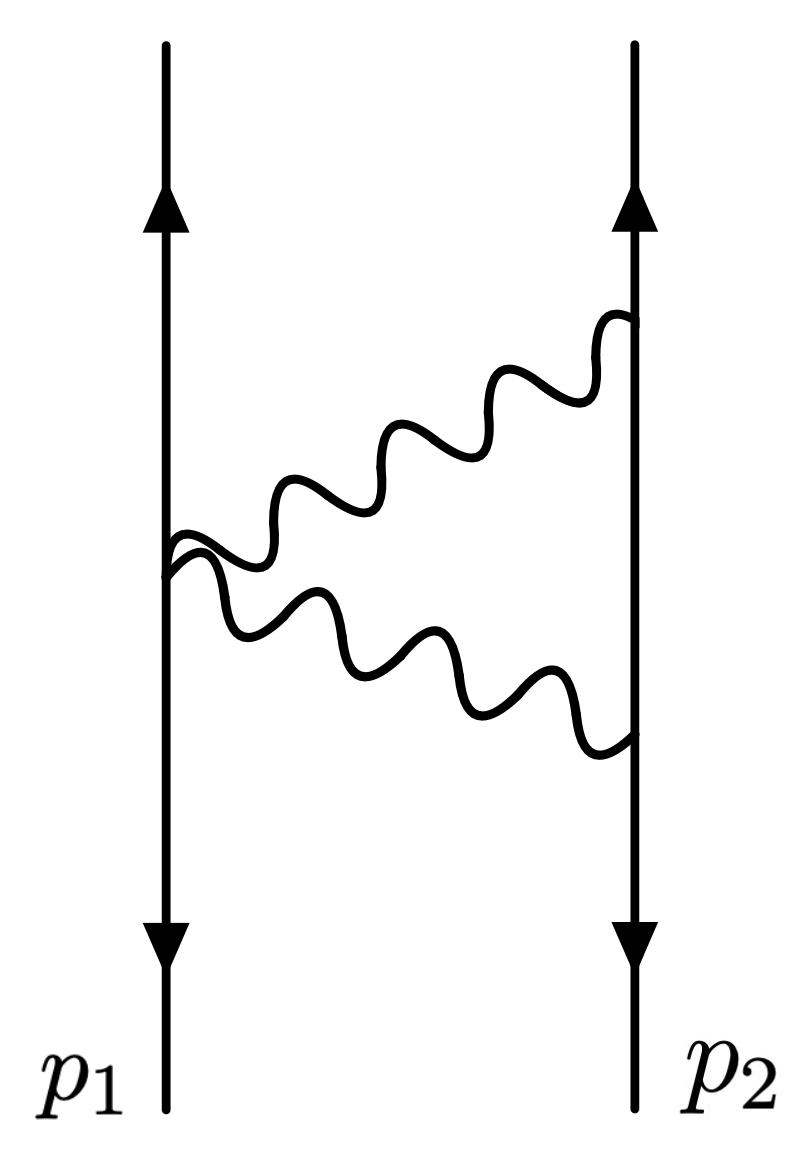}
            \label{subfig:B}
        }
        \caption{The two diagrams containing classical information for the $2\rightarrow 2$ elastic scattering at 2PM.}
        \label{fig1}
    \end{figure}

In order to get the classical 2PM amplitude we consider the following cut,
\begin{equation}
\label{F65}
i\mathcal{M}^{2\mathrm{PM}}(a;p_i,q)=
\vcenter{\hbox{\includegraphics[width=6.6cm,height=4.6cm]{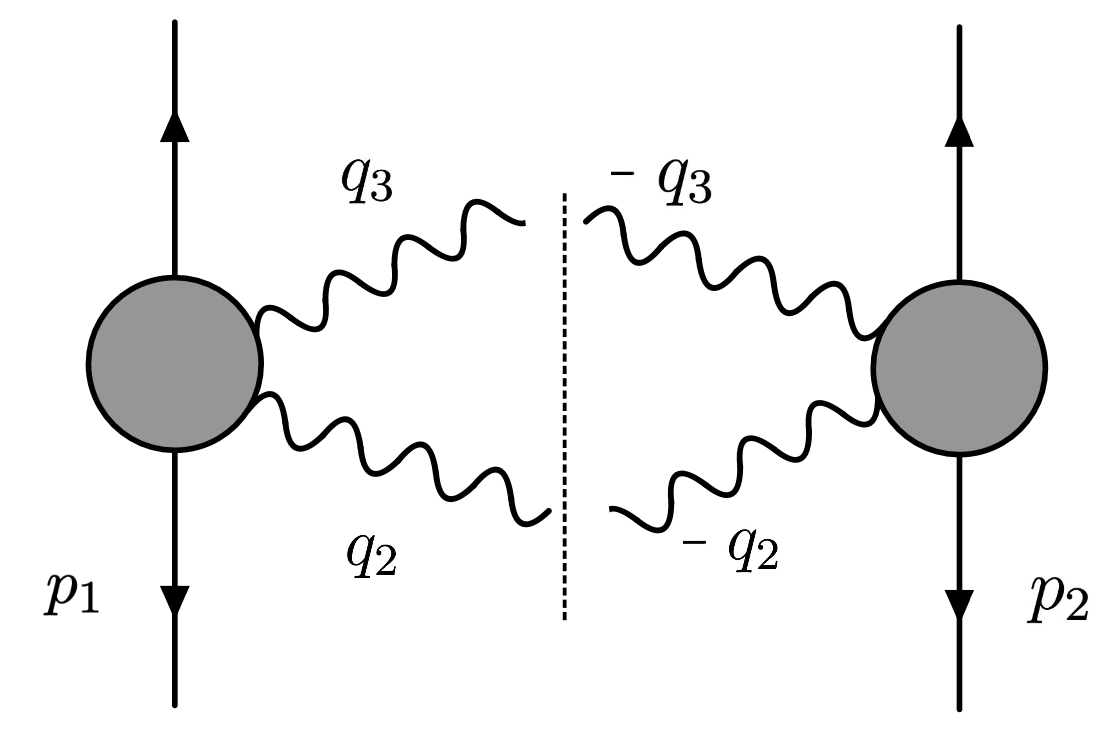}}}=\int\frac{d^4l}{(2\pi)^4}M^{2\mathrm{PM}}(a;p_i,q_i),
\end{equation}
where $l$ is the loop momentum and can be chosen to be either $q_2$ or $q_3$. Explicitly, the one-loop integrand is,
\begin{align}
\label{F65.1}
\nonumber M^{2\mathrm{PM}}(a;p_i,q_i)=&\mathcal{M}^{\mu_1\nu_1;\rho_1\sigma_1}_4(a;p_1,-q-q_3,q_3)[P(q_3)]_{\mu_1\nu_1;\mu_2\nu_2}[P(q_3+q)]_{\rho_1\sigma_1;\rho_2\sigma_2}\\&\mathcal{M}^{\mu_2\nu_2;\rho_2\sigma_2}_4(a;p_2,-q-q_3,-q_3),
\end{align}
where $q=-q_2-q_3$ is the transfer momentum and where $[P(q)]_{\mu\nu;\rho\sigma}$ is the de Donder propagator,
\begin{align}
\label{F65.2}
[P(q)]_{\mu\nu;\rho\sigma}=-\frac{i}{q^2}\Pi^{(h)}_{\mu\nu;\rho\sigma}.
\end{align}
Instead of computing the product in \eqref{F65.1}, we find it convenient to switch to on-shell variables here. In terms of spinor-helicity variables, the Compton amplitudes in \eqref{F61} are given by,
\begin{align}
\label{F64}
&\mathcal{M}^{++}_4(a;p_i,q_i)=\frac{\kappa^2}{4} e^{-a\cdot (q_2+q_3)}\frac{m^4[23]^4}{t\tau_{12}\tau_{13}},\quad &&\mathcal{M}^{--}_4(a;p_i,q_i)=\frac{\kappa^2}{4}e^{a\cdot (q_2+q_3)}\frac{m^4\langle 23\rangle^4}{t\tau_{12}\tau_{13}},\\\nonumber\\&\mathcal{M}^{+-}_4(a;p_i,q_i)=\frac{\kappa^2}{4}e^{-a\cdot (q_2-q_3)}\frac{[ 2|p_1|3\rangle ^4}{t\tau_{12}\tau_{13}},\quad &&\mathcal{M}^{-+}_4(a;p_i,q_i)=\frac{\kappa^2}{4}e^{a\cdot (q_2-q_3)}\frac{[ 3|p_1|2\rangle^4}{t\tau_{12}\tau_{13}},
\end{align}
where $\tau_{ij}=2 p_i\cdot q_j$, $t=s_{23}=2q_2\cdot q_3=q^2$. Following similar considerations \textit{e.g.} as in \cite{Bjerrum-Bohr:2013bxa,Bern:2019crd}, the full  integrand of the classical one-loop amplitude can easily be recast in the following form,
\begin{align}
\label{F66}
\nonumber M^{2\mathrm{PM}}(a;p_i,q_i)&=\frac{\kappa^4}{16}\frac{1}{t\tau_{12}\tau_{13}\tau_{22}\tau_{23}q_2^2q_3^2}\bigg\{2t^4m_1^4m_2^4\cosh(a\cdot(q_3+q_2))\\\nonumber &+\cosh(a\cdot(q_3-q_2))[\mathrm{tr}_+(q_2p_2q_3p_1)^4+\mathrm{tr}_-(q_2p_2q_3p_1)^4]\\&+\sinh(a\cdot(q_3-q_2))[\mathrm{tr}_+(q_2p_2q_3p_1)^4-\mathrm{tr}_-(q_2p_2q_3p_1)^4]\bigg\},
\end{align}
where the chiral traces are,
\begin{align}
\label{F67}
\nonumber\mathrm{tr}_{\pm}(q_2p_2q_3p_1)&=\frac{1}{2}\tau_{22}\tau_{13}+\frac{1}{2}\tau_{23}\tau_{12}-\frac{t}{2}(s-m_1^2-m_2^2)\pm 2i\epsilon(q_2,p_2,q_3,p_1)\\&\equiv \mathcal{E}\pm\mathcal{O},
\end{align}
where $s=(p_1+p_2)^2$ and where we introduced the notation \cite{Bern:2019crd}, 
\begin{align}
\label{F68}
\mathcal{E}= \frac{1}{2}\tau_{22}\tau_{13}+\frac{1}{2}\tau_{23}\tau_{12}-\frac{t}{2}(s-m_1^2-m_2^2),\qquad \mathcal{O}=2i\epsilon(q_2,p_2,q_3,p_1).
\end{align}
Squaring $\mathcal{O}$ and computing the Gram determinant gives,
\begin{align}
\label{F69}
\mathcal{O}^2=-4\epsilon(q_2,p_2,q_3,p_1)^2=\mathcal{E}^2-(m_1^2t-\tau_{12}\tau_{13})(m_2^2t-\tau_{22}\tau_{23}).
\end{align}
Using $2p_1\cdot q=q^2$ and $2p_2\cdot q=-q^2$ we get,
\begin{align}
\label{F71}
\frac{1}{\tau_{13}}+\frac{1}{\tau_{12}}=-\frac{t}{\tau_{13}\tau_{12}},\qquad\frac{1}{\tau_{23}}+\frac{1}{\tau_{22}}=\frac{t}{\tau_{23}\tau_{22}},
\end{align}
that, inserted into \eqref{F66} yields
for $M^{2\mathrm{PM}}(a)$,
\begin{align}
\label{F70}
\nonumber &M^{2\mathrm{PM}}(a;p_i,q_i)=-\frac{\kappa^4}{16}\frac{1}{q_2^2q_3^2}\bigg[f_{\mathrm{even}}(\tau_{12},\tau_{13},\tau_{22},\tau_{23})\cosh(a\cdot (q_3-q_2))\\\nonumber&+f_{\mathrm{odd}}(\tau_{12},\tau_{13},\tau_{22},\tau_{23})\sinh(a\cdot (q_3-q_2))\bigg]\bigg(\frac{1}{\tau_{13}\tau_{23}}+\frac{1}{\tau_{13}\tau_{22}}+\frac{1}{\tau_{12}\tau_{23}}\\&+\frac{1}{\tau_{12}\tau_{22}}\bigg),
\end{align}
where we neglected the first term in \eqref{F66} because it corresponds to a ``box" diagram whose classical limit, in $D=4$, is compensated by the Born subtraction of the effective theory (see \textit{e.g.} \cite{Cristofoli:2020uzm}) and where we defined,
\begin{align}
\label{F72}
&f_{\mathrm{even}}(\tau_{12},\tau_{13},\tau_{22},\tau_{23})\equiv\frac{1}{t^4}(2\mathcal{E}^4+2\mathcal{O}^4+12\mathcal{E}^2\mathcal{O}^2),\\ &f_{\mathrm{odd}}(\tau_{12},\tau_{13},\tau_{22},\tau_{23})\equiv\frac{1}{t^4}(8\mathcal{E}^3\mathcal{O}+8\mathcal{E}\mathcal{O}^3),
\end{align}
to be the even and odd contributions in the spin multipole expansion, respectively. We start by analysing the even part. We notice that the variables $\tau_{ij}$ are not all independent and infact they are related by,
\begin{align}
\label{F73}
\tau_{22}=2p_2\cdot q_2=t-\tau_{23},\qquad \tau_{12}=2p_1\cdot q_2=-t-\tau_{13}.
\end{align}
This enables us to define the functions,
\begin{align}
\label{F74}
&{}_{(1)}f_{\mathrm{even}}(\tau_{13},\tau_{23})=f_{\mathrm{even}}(-t-\tau_{13},\tau_{13},t-\tau_{23},\tau_{23})=\sum_{i,j=0}^4{}_{(1)}f_{\mathrm{even}}^{ij}(\tau_{13})^i(\tau_{23})^j,\\\nonumber\\&{}_{(2)}f_{\mathrm{even}}(\tau_{13},\tau_{22})=f_{\mathrm{even}}(-t-\tau_{13},\tau_{13},\tau_{22},t-\tau_{22})=\sum_{i,j=0}^4{}_{(2)}f_{\mathrm{even}}^{ij}(\tau_{13})^i(\tau_{22})^j,\\\nonumber\\&{}_{(3)}f_{\mathrm{even}}(\tau_{12},\tau_{23})=f_{\mathrm{even}}(\tau_{12},-t-\tau_{12},t-\tau_{23},\tau_{23})=\sum_{i,j=0}^4{}_{(3)}f_{\mathrm{even}}^{ij}(\tau_{12})^i(\tau_{23})^j,\\\nonumber\\&{}_{(4)}f_{\mathrm{even}}(\tau_{12},\tau_{22})=f_{\mathrm{even}}(\tau_{12},-t-\tau_{12},\tau_{22},t-\tau_{22})=\sum_{i,j=0}^4{}_{(4)}f_{\mathrm{even}}^{ij}(\tau_{12})^i(\tau_{22})^j.
\end{align}
Inserting the above expansions in equation \eqref{F70} will give terms having either two massless propagators, the so-called ``bubbles", or three propagators of which two massless and one massive, the so called ``triangles". Only the latter contribute in the classical limit and we now proceed to extract them. We have,
\begin{align}
\label{F75}
\nonumber M^{2\mathrm{PM}}_{\mathrm{even}}(a;p_i,q_i)=&-\frac{\kappa^4}{16}\frac{1}{q_2^2q_3^2}\sum_{i=0}^3\bigg[\frac{{}_{(1)}f_{\mathrm{even}}^{0,i+1}(\tau_{23})^i+{}_{(2)}f_{\mathrm{even}}^{0,i+1}(\tau_{22})^i}{\tau_{13}}\\\nonumber &+\frac{{}_{(3)}f_{\mathrm{even}}^{0,i+1}(\tau_{23})^i+{}_{(4)}f_{\mathrm{even}}^{0,i+1}(\tau_{22})^i}{\tau_{12}}+\frac{{}_{(1)}f_{\mathrm{even}}^{i+1,0}(\tau_{13})^i+{}_{(3)}f_{\mathrm{even}}^{i+1,0}(\tau_{12})^i}{\tau_{23}}+\\\nonumber&+\frac{{}_{(2)}f_{\mathrm{even}}^{i+1,0}(\tau_{13})^i+{}_{(4)}f_{\mathrm{even}}^{i+1,0}(\tau_{12})^i}{\tau_{22}}\bigg]\cosh(a\cdot (q_3-q_2))\\&\equiv {}_{\triangleright}M^{2\mathrm{PM}}_{\mathrm{even}}(a;p_i,q_i)+{}_{\triangleleft}M^{2\mathrm{PM}}_{\mathrm{even}}(a;p_i,q_i),
\end{align}
where we included in the $\triangleright$ and $\triangleleft$ contributions diagrams having triangular topology with one massive propagator with mass $m_1$ and $m_2$, respectively, as shown in Figure \ref{fig1}.\\
We choose as independent loop momentum $l=q_3$ and therefore we substitute in \eqref{F75} $q_2=-l-q$ and $\tau_{12}=-t-\tau_{13}$ and $\tau_{22}=t-\tau_{23}$. Taking the classical $t\rightarrow 0$ limit yields,
\begin{align}
\label{F76}
\mathcal{M}^{2\mathrm{PM}}_{\mathrm{even}}(a;p_i,q)=i\frac{\kappa^4}{16}\bigg[c_{\triangleright}\mathcal{I}_{\triangleright}(a)+(c_{\triangleright})_{\mu\nu}\mathcal{I}^{\mu\nu}_{\triangleright}(a)+(\triangleright\leftrightarrow \triangleleft)\bigg],
\end{align}
where,
\begin{align}
\label{F77}
c_{\triangleright}=4m_1^2[m_1^4+m_1^2(m_2^2-2s)+(m_2^2-s)^2],\qquad (c_{\triangleright})_{\mu\nu}=-\frac{8m_1^4p_{2\mu}p_{2\nu}}{t},
\end{align}
and,
\begin{align}
\label{F78}
\mathcal{I}^{\mu_1\dots\mu_n}_{\triangleright}(a)=\int\frac{d^4l}{(2\pi)^4}\frac{l^{\mu_1}\dots l^{\mu_n}\cosh(2a\cdot l+a\cdot q)}{l^2(l+q)^2[(p_1+l)^2-m_1^2]},
\end{align}
and where quantities with $\triangleleft$ can be obtained from $\triangleright$ by sending $m_1\leftrightarrow m_2$ and $p_1\leftrightarrow p_2$.

Since we are ultimately interested in obtaining the scattering angle, from now on we consider an aligned spin configuration, where the spins of the black holes are aligned to the $z$ axis and perpendicular to the scattering plane $xy$. We also choose to work in the center of mass frame and we fix,
\begin{align}
\label{F79}
p_1=(-E_1,-p,0,0),\qquad p_2=(-E_2,p,0,0),\qquad a_i=(0,0,0,a_i).
\end{align}
Using \eqref{FB9} and \eqref{FB9.3}, we find in the classical limit,
\begin{align}
\label{F80}
\nonumber \mathcal{M}^{2\mathrm{PM}}_{\mathrm{even}}(a;p_i,q)=&\frac{\kappa^4}{16}\frac{1}{32\sqrt{-q^2}}\bigg\{4m_1[m_1^4+m_1^2(m_2^2-2s)+(m_2^2-s)^2]I_0(\mathscr{F}(a,q))\\&+2m_1\bigg[m_1^2m_2^2-\frac{(s-m_1^2-m_2^2)^2}{4}\bigg]\frac{I_1(\mathscr{F}(a,q))}{\mathscr{F}(a,q)}+(m_1\leftrightarrow m_2)\bigg\},
\end{align}
where $I_0(x)$ and $I_1(x)$ are modified Bessel functions of the first kind and their argument is, \begin{align}
\label{F80.5}
\mathscr{F}(a,q)=\sqrt{(a\cdot q)^2-a^2q^2},
\end{align}
which is shift symmetric. We checked that equation \eqref{F80} reduces to the standard 2PM result for $a=0$, \textit{e.g.} \cite{Cristofoli:2020uzm}. Let us now consider the part of the 2PM amplitude containing odd powers of the classical spin vector. We get,
\begin{align}
\label{F81}
\mathcal{M}^{2\mathrm{PM}}_{\mathrm{odd}}(a;p_i,q)=\frac{\kappa^4}{16}\bigg[d_{\triangleright}\epsilon_{\mu}(q,p_2,p_1)\mathcal{J}^{\mu}_{\triangleright}(a)+(d_{\triangleright})_{\nu}\epsilon_{\mu}(q,p_2,p_1)\mathcal{J}^{\mu\nu}_{\triangleright}(a)+(\triangleright\leftrightarrow \triangleleft)\bigg],
\end{align}
where,
\begin{align}
\label{F82}
d_{\triangleright}=\frac{8m_1^2(m_1^2+m_2^2-s)}{t},\qquad (d_{\triangleright})_{\nu}=-\frac{16m_1^2p_{2\nu}}{t},
\end{align}
and,
\begin{align}
\label{F83}
\mathcal{J}^{\mu_1\dots\mu_n}(a)=\int\frac{d^4l}{(2\pi)^4}\frac{l^{\mu_1}\dots l^{\mu_n}\sinh(2a\cdot l+a\cdot q)}{l^2(l+q)^2[(l+p_1)^2-m_1^2]}.
\end{align}
We find that in the classical limit only the vector integral contributes in \eqref{F81} and we have, using \eqref{FB10.1},
\begin{align}
\label{F83}
\nonumber\mathcal{M}^{2\mathrm{PM}}_{\mathrm{odd}}(a;p_i,q)=&\frac{\kappa^4}{16}\bigg[\frac{m_1(m_1^2+m_2^2-s)}{32\sqrt{-q^2}}i\epsilon(a,q,p_2,p_1)\frac{I_1(\mathscr{F}(a,q))}{\mathscr{F}(a,q)}\\&+(m_1\leftrightarrow m_2)\bigg].
\end{align}
Computing the Gram determinant gives (see \eqref{FA4.1}-\eqref{FA4.2}),
\begin{align}
\label{F85}
\epsilon(a,q,p_2,p_1)=i\mathscr{F}(a,q)m_1m_2\sqrt{\sigma^2-1}+\mathcal{O}(t^2).
\end{align}
Substituting into \eqref{F83} gives the final result for the classical 2PM amplitude odd in the spin,
\begin{align}
\label{F86}
\nonumber\mathcal{M}^{2\mathrm{PM}}_{\mathrm{odd}}(a;p_i,q)=&-\frac{\kappa^4}{16}\frac{\sqrt{\sigma^2-1}}{32\sqrt{-q^2}}\bigg[m_1(m_1^2+m_2^2-s)I_1(\mathscr{F}(a,q))\\&+(m_1\leftrightarrow m_2)\bigg],
\end{align}
where the Lorentz factor is related to $s$ by $\sigma=\frac{1}{2m_1m_2}(s-m_1^2-m_2^2)$.

The 2PM eikonal phase is,
\begin{align}
\label{F87}
\nonumber\tilde{\mathcal{M}}^{2\mathrm{PM}}(\mathbf{a};p_i,\mathbf{b})&=\frac{1}{4m_1m_2\sqrt{\sigma^2-1}}\int\frac{d^2\mathbf{q}}{(2\pi)^2}\mathcal{M}^{2\mathrm{PM}}(\mathbf{a};p_i,\mathbf{q})e^{-i\mathbf{q}\cdot\mathbf{b}}\\\nonumber\\&=\tilde{\mathcal{M}}^{2\mathrm{PM}}_{\mathrm{even}}(\mathbf{a};p_i,\mathbf{b})+\tilde{\mathcal{M}}_{\mathrm{odd}}^{2\mathrm{PM}}(\mathbf{a};p_i,\mathbf{b}).
\end{align}
where,
\begin{align}
\label{F88}
\nonumber \tilde{\mathcal{M}}^{2\mathrm{PM}}_{\mathrm{even}}(\mathbf{a};p_i,\mathbf{b})=&\frac{\kappa^4}{16}\frac{1}{128m_1m_2\sqrt{\sigma^2-1}}\bigg\{4m_1[m_1^4+m_1^2(m_2^2-2s)+(m_2^2-s)^2]\tilde{I}_0[\mathbf{a},\mathbf{b}]\\&+2m_1\bigg[m_1^2m_2^2-\frac{(s-m_1^2-m_2^2)^2}{4}\bigg]\tilde{\bigg(\frac{I_1}{\mathscr{F}}\bigg)}[\mathbf{a},\mathbf{b}]+(m_1\leftrightarrow m_2)\bigg\},
\end{align}
and,
\begin{align}
\label{F89}
\nonumber \tilde{\mathcal{M}}^{2\mathrm{PM}}_{\mathrm{odd}}(\mathbf{a};p_i,\mathbf{b})=&-\frac{\kappa^4}{16}\frac{1}{128m_1m_2}\bigg[m_1(m_1^2+m_2^2-s)\tilde{I}_1[\mathbf{a},\mathbf{b}]\\&+(m_1\leftrightarrow m_2)\bigg],
\end{align}
where the Fourier transforms of the Bessel functions are computed in appendix \ref{AC} in \eqref{FC4}, \eqref{FC5} and \eqref{FC6}. The 2PM scattering angle is defined through the usual relation,
\begin{align}
\label{F90}
\Theta^{2\mathrm{PM}}(a;p,b)=-\frac{1}{p}\frac{\partial}{\partial b}\mathcal{M}^{\mathrm{2PM}}(a;p,b)=\Theta^{2\mathrm{PM}}_{\mathrm{even}}(a;p,b)+\Theta^{2\mathrm{PM}}_{\mathrm{odd}}(a;p,b),
\end{align}
where,
\begin{align}
\label{F91}
\Theta^{2\mathrm{PM}}_{\mathrm{even}}(a;p,b)=-\frac{1}{p}\frac{\partial}{\partial b}\mathcal{M}^{\mathrm{2PM}}_{\mathrm{even}}(a;p,b),\qquad \Theta^{2\mathrm{PM}}_{\mathrm{odd}}(a;p,b)=-\frac{1}{p}\frac{\partial}{\partial b}\mathcal{M}^{\mathrm{2PM}}_{\mathrm{odd}}(a;p,b).
\end{align}
\section{Conclusions and outlook}
We introduced a new Yang-Mills type of Lagrangian based on a notion of minimal coupling that incorporates classical spin effects. In particular, we found that the three-point amplitude of $\sqrt{\mathrm{Kerr}}$ can consistently be described using a ``classical spin covariant derivative", whose implementation at the level of the Lagrangian automatically yields contact terms. We used such theory to compute the Compton amplitude to all orders in the spin, first in gauge theory and then, using classical double copy, in gravity. Such amplitudes, coming directly from a local Lagrangian, are not affected by spurious poles and they reproduce the  results of \cite{Arkani-Hamed:2017jhn,Bautista:2019tdr,Saketh:2022wap,Bautista:2022wjf,Kim:2023drc} in the equal-helicity sector whereas they do not in the opposite-helicity one. In particular, we showed that the opposite-helicity Compton amplitude coming from our model does not depend on the complex vector $w$ present  \textit{e.g.} in \cite{Aoude:2020onz,Saketh:2022wap}. Including $w$ as shown in section \ref{S4} yields results that agree, up to $a^4$, with the ones presented in the above mentioned references. We concluded with a 1-loop (2PM) computation of the scattering amplitude for an elastic $2\rightarrow 2$ process in an aligned spin configuration, to all orders in the classical spin vector, from which we extracted the 2PM eikonal and scattering angle. 

As remarked above, the field theory constructed here does not capture all the features characterizing $\sqrt{\mathrm{Kerr}}$ and Kerr solutions and their interactions with the gauge fields. However, it has to be regarded as a minimal model to which one is free to add gauge invariant terms of higher order in the curvature in order to reproduce observables relevant for the dynamics of true Kerr binaries in the PM regime. The main advantage of this theory relies in its intrinsic simplicity, which we believe to be a fundamental criterion for the description of Kerr black holes as elementary particles. Indeed, the 1PM and the 2PM elastic scattering amplitudes have a very simple mathematical structure. Concerning future directions, it would be interesting to go further in the PM expansion and to explicitly compute the 3PM scattering amplitude and eikonal, both in the conservative and radiative sectors \footnote{Note that the result contained in \cite{Alessio:2022kwv} for the radiative sector at 3PM, relying only on the three-point amplitude as main building block, is fully consistent with the theory presented here.} as well as to construct a worldline version of this theory.

\label{S7}
\subsection*{Acknowledgements} 
I would like to thank Maor Ben-Shahar, Zvi Bern, Luca Buoninfante, Paolo Di Vecchia, Alessandro Georgoudis, Kays Haddad, Carlo Heissenberg, Gustav Jakobsen, Henrik Johansson, Gustav Mogull, Stavros Mougiakakos, Alexander Ochirov, Paolo Pichini, Jan Plefka and Justin Vines for very useful discussions. I thank Henrik Johansson and Paolo Pichini for their interest during initial stages of the project. I am especially grateful to Kays Haddad, Henrik Johansson and Paolo Pichini for enlightening discussions and to Paolo Di Vecchia and Paolo Pichini for comments on this draft. The research of FA is fully supported by the Knut and Alice Wallenberg Foundation under grant KAW 2018.0116. Nordita is partially supported by Nordforsk.
\appendix
\section{Kerr-Schild coordinates and Newman-Janis shift}
\label{S2}
In this appendix we review some features of the Schwarzschild and Kerr solutions using the Kerr-Schild ansatz \cite{Kerr:1963ud,Kerr2,Debney:1969zz}, that consists in parametrizing the full spacetime metric $g_{\mu\nu}$ as,
\begin{align}
\label{F1}
g_{\mu\nu}=\bar{g}_{\mu\nu}+\hspace{0.05cm}\phi\hspace{0.05cm} k_{\mu}k_{\nu},\hspace{1cm}\bar{g}_{\mu\nu}{k}^{\mu}k^{\nu}=0=g_{\mu\nu}{k}^{\mu}k^{\nu},
\end{align}
where $\bar{g}_{\mu\nu}$ is a background solution satisfying $\bar{R}_{\mu\nu}=0$. Because $k^{\mu}$ is null, its indices can be raised and lowered with both $g_{\mu\nu}$ and $\bar{g}_{\mu\nu}$. Vacuum Einstein equations $R_{\mu\nu}=0$, together with \eqref{F1} imply \cite{Lee:2018gxc},
\begin{align}
\label{F2}
R^{(1)}_{\mu\nu}\equiv \bar{\nabla}_{\rho}\bigg(\bar{\nabla}_{(\mu}(\phi\hspace{0.05cm} k_{\nu)}k^{\rho})-\frac{1}{2}\bar{\nabla}^{\rho}(\phi \hspace{0.05cm}k_{\mu}k_{\nu})\bigg)=0.
\end{align}
We are interested in describing black holes and hence, from now on we choose $\phi$ to depend only on the radial coordinate $r$ and on the coordinate $z$, \textit{i.e.} we specialize to axisymmetric solutions. Furthermore, we fix $k^{\mu}=(1,-1,0,0)$ \cite{Arkani-Hamed:2019ymq} and choose $\bar{g}_{\mu\nu}$ to be the flat Minkowski metric $\eta_{\mu\nu}$.

If we work in spherical coordinates $(r,\theta,\phi)$ defined by $x+i y=re^{i\phi}\sin\theta $ and $z=r\cos\theta$, it is immediate to prove that $R^{(1)}_{\mu\nu}=0$ implies,
\begin{align}
\label{F3}
\phi+r\partial_{r}\phi=0,\hspace{1cm}\partial_r\partial_{\theta}\phi=0,
\end{align}
whose solution is,
\begin{align}
\label{F4}
\phi_S=\frac{r_s}{r},\hspace{1cm},
\end{align}
where $r_s=2G_N m$, that is precisely the Schwarzschild black hole in Kerr-Schild coordinates.

 To obtain the Kerr solution, we perform a change of coordinates and parametrize $\eta_{\mu\nu}$ with oblate spheroidal coordinates, defined by $x+iy=(r+ia)e^{i\phi}\sin\theta$ and $z=r\cos\theta$. Notice that so far, the parameter $a$ is not related to the spin and it appears as mere parameter characterizing the foliation of the spatial part of the spacetime metric. Equation \eqref{F2} reduces to,
\begin{align}
\label{F5}
\frac{r^2-a^2\cos^2\theta}{r^2+a^2\cos^2\theta}\phi+r\partial_r\phi=0,\hspace{1cm}\partial_r\partial_{\theta}\phi-\partial_r\bigg(\frac{2a^2\phi\sin\theta\cos\theta}{r^2+a^2\cos^2\theta}\bigg)=0,
\end{align}
solved by,
\begin{align}
\label{F6}
\phi_K=\frac{r_s r}{r^2+a^2\cos^2\theta},
\end{align}
which is the Kerr-Schild form of Kerr spacetime. We note here that performing a change of coordinates from spherical to oblate spheroidal on the Schwarzschild solution \eqref{F4} \textit{does not} yield Kerr. Indeed, the two black holes are not diffeomorphic. 
Only performing the diffeomorphism on the ansatz for the metric, \textit{before} solving Einstein equations, gives Kerr spacetime. One of the advantages of using Kerr-Schild coordinates is that they emphasize the similarities between Schwarzschild and Kerr and that there is a simple and clear connection between them called Newman-Janis shift \cite{Newman:1965tw}. It consists in temporarily complexifying the radial coordinate of the Schwarzschild solution and then perform the complex translation $r\rightarrow r+ia\cos\theta$ to reconstruct Kerr spacetime as follows:
\begin{align}
\label{F7}
\phi_K=\frac{1}{2}(\phi_S+\phi_S^*)|_{r\rightarrow r+ia\cos\theta}=\frac{r_s}{2}\bigg(\frac{1}{r+ia\cos\theta}+\frac{1}{r-ia\cos\theta}\bigg).
\end{align}
This equation displays that all the information about the spinning solution are actually encoded into the spinless one. All the curvature invariants constructed with the Kerr metric can be obtained by Newman-Janis shifting the Schwarzschild ones, \textit{e.g.}
\begin{align}
\label{F8}
R^S_{\mu\nu\rho\sigma}R^{\mu\nu\rho\sigma}_S=\frac{48 G^2 m^2}{r^6},
\end{align}
and,
\begin{align}
\label{F9}
R^K_{\mu\nu\rho\sigma}R^{\mu\nu\rho\sigma}_K=24G^2m^2\frac{2r^6-30a^2r^4\cos^2\theta+30a^4r^2\cos^4\theta-2a^6\cos^6\theta}{(r^2+a^2\cos^2\theta)^6},
\end{align}
are simply related by,
\begin{align}
\label{F10}
R^K_{\mu\nu\rho\sigma}R^{\mu\nu\rho\sigma}_K=\frac{1}{2}(R^S_{\mu\nu\rho\sigma}R^{\mu\nu\rho\sigma}_S+(R^{S}_{\mu\nu\rho\sigma}R^{\mu\nu\rho\sigma}_S)^*)|_{r\rightarrow r+ia\cos\theta}.
\end{align}
The construction of the field theory described in the paper is based on an implementation of the Newman-Janis shift at the level of the Lagrangian.

We end this section by showing the gauge theory, single copy classical solutions corresponding to Schwarzschild and Kerr, namely Coulomb and $\sqrt{\mathrm{Kerr}}$. We parametrize the gauge field as,
\begin{align}
\label{F11}
A^{\mu}=\phi k^{\mu},\hspace{1cm} \eta_{\mu\nu}k^{\mu}k^{\nu}=0.
\end{align}
with $k^{\mu}=(1,-1,0,0)$. In spherical and oblate spheroidal coordinates, the solutions to equations of motion $D_{\mu}F^{\mu\nu}=0$ for $\phi$ turn out to be,
\begin{align}
\label{F12}
A^{\mu}_S=\phi_Sk^{\mu},\hspace{1cm} A^{\mu}_K=\phi_Kk^{\mu},
\end{align}
respectively, where again $\phi_S$ and $\phi_K$ are given in \eqref{F4} and \eqref{F6} with the replacement $r_s\rightarrow g$. As discussed in \cite{Monteiro:2014cda,Monteiro:2020plf,Monteiro:2021ztt,Lee:2018gxc,Arkani-Hamed:2019ymq}, the classical double copy structure between the gauge theory and gravity solutions is evident. 
\section{Parametrization for the momenta}
\label{AC1}
The parametrization for the momenta and the polarization vectors we use at the end of section \ref{S4} is,
\begin{align}
\label{F40}
&p_1=-(m,0,0,0),\qquad &&q_2=-\hbar(\omega,0,0,\omega),\\
\label{F41}
&q_3=\frac{\hbar\omega}{1+\frac{2\hbar\omega}{m}\sin^2(\theta/2)}(1,0,\sin\theta,\cos\theta),\qquad &&p_4=-p_1-q_2-q_3,\\
\label{F42}
&\epsilon_{2}^{+}=\frac{1}{\sqrt{2}}(0,1,i,0),\qquad &&\epsilon_{2}^{-}=-\frac{1}{\sqrt{2}}(0,1,-i,0),\\\label{F43.1}&\epsilon_{3}^{+}=\frac{1}{\sqrt{2}}(0,\cos\theta,i,-\sin\theta),\qquad &&\epsilon_{3}^{-}=-\frac{1}{\sqrt{2}}(0,\cos\theta,-i,-\sin\theta).
\end{align}
\section{1PM amplitude for $\sqrt{\mathrm{Kerr}}$}
\label{AB}
The 1PM amplitude between two $\sqrt{\mathrm{Kerr}}$ particles having charges $e_1$ and $e_2$ exchanging one photon with momentum $q$ can be obtained by gluing together two three-point amplitudes with the Feynman propagator, as shown in \eqref{FA1},
\begin{equation}
\label{FA1}
i\mathcal{A}^{1\mathrm{PM}}(a;p_i,q)=\vcenter{\hbox{\includegraphics[width=4.5cm,height=4.4cm]{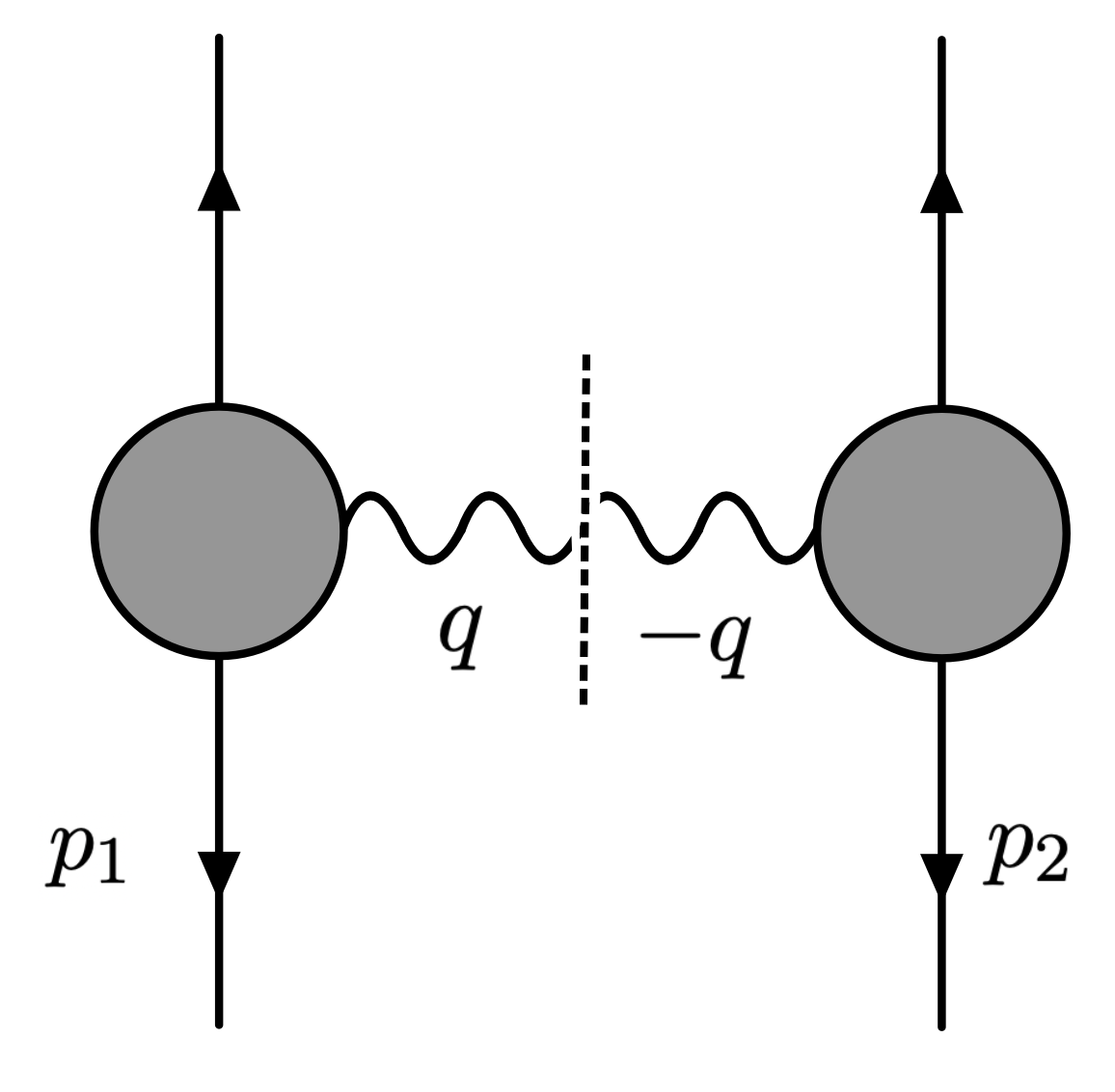}}},
\end{equation}
because in the classical limit only the $t=q^2$ pole contributes. Before doing so, we use three-point kinematics to rewrite $\mathcal{A}^{\mu}_3(a,p)$ in a more convenient form,
\begin{align}
\label{FA2}
\nonumber \mathcal{A}^{\mu}_3(a;p,q)&=-ie\hspace{0.03cm}\mathrm{exp}\{i\epsilon^{\mu}{}_{\rho}(a,q)\}p^{\rho}\\&=-ie\left(p^{\mu}\cosh(a\cdot q)+i\frac{\epsilon^{\mu}(p,a,q)}{a\cdot q}\sinh(a\cdot q)\right).
\end{align}
Therefore, the 1PM amplitude is given by,
\begin{align}
\label{FA3}
\nonumber i\mathcal{A}^{1\mathrm{PM}}(a;p_i,q)=&-\frac{i}{q^2}\eta_{\mu\nu}\mathcal{A}_3^{\mu}(a_1;p_1,q)\mathcal{A}^{\nu}_3(a_2,p_2,-q)\\\nonumber =&\frac{i}{q^2}e_1 e_2\bigg(p_1\cdot p_2\cosh(a_1\cdot q)\cosh(a_2\cdot q)\\\nonumber&+\epsilon^{\mu}(p_1,a_1,q)\epsilon_{\mu}(p_2,a_2,q)\frac{\sinh(a_1\cdot q)\sinh(a_2\cdot q)}{a_1\cdot q a_2\cdot q}\\\nonumber&+i\epsilon(p_2,p_1,a_1,q)\cosh(a_2\cdot q)\frac{\sinh(a_1\cdot q) }{a_1\cdot q}\\&-i\epsilon(p_1,p_2,a_2,q)\cosh(a_1\cdot q)\frac{\sinh(a_2\cdot q) }{a_2\cdot q}\bigg).
\end{align}
Using $q\cdot p_1=0=q\cdot p_2$ and $p_1\cdot a_1=0=p_2\cdot a_2$, together with $q^2=0$, we compute the following Gram determinants \cite{Guevara:2019fsj},
\begin{align}
\label{FA4}
&\epsilon^{\mu}(p_1,a_1,q)\epsilon_{\mu}(p_2,a_2,q)=m_1m_2\sigma a_1\cdot q a_2\cdot q,\\
\label{FA4.1}
&\epsilon(p_1,p_2,a_1,q)=im_1m_2(a_1\cdot q)\sqrt{\sigma^2-1},\\
\label{FA4.2} &\epsilon(p_1,p_2,a_2,q)=im_1m_2(a_2\cdot q)\sqrt{\sigma^2-1},
\end{align}
where $\sigma=\frac{p_1\cdot p_2}{m_1m_2}$ is the Lorentz factor. Substituting the above relations back into \eqref{FA3} we get,
\begin{align}
\label{FA5}
\nonumber i\mathcal{A}^{1\mathrm{PM}}(a;p_i,q)=&\frac{ie_1 e_2 m_1 m_2}{q^2}\bigg\{\sigma\bigg[\cosh\left(a_1\cdot q\right)\cosh\left(a_2\cdot q\right)+\sinh\left(a_1\cdot q\right)\sinh\left(a_2\cdot q\right)\bigg]\\\nonumber &+\sqrt{\sigma^2-1}\bigg[\cosh\left(a_2\cdot q\right)\sinh\left(a_1\cdot q\right)+\cosh\left(a_1\cdot q\right)\sinh\left(a_2\cdot q\right)\bigg]\bigg\}\\  =&\frac{ie_1e_2m_1m_2}{q^2}\bigg[\sigma\cosh(a\cdot q)+\sqrt{\sigma^2-1}\sinh(a\cdot q)\bigg],
\end{align}
where $a=a_1+a_2$. Introducing the relative velocity of the two bodies $v=\frac{\sqrt{\sigma^2-1}}{\sigma}$ we get for the classical 1PM amplitude,
\begin{align}
\label{FA6}
\mathcal{A}^{1\mathrm{PM}}(a;p_i,q)=\frac{ e_1m_1 e_2m_2}{2q^2}\sigma\sum_{\pm}(1\pm v)e^{\pm i\frac{\epsilon(p_1,p_2,q,a)}{m_1m_2\sigma v}},
\end{align}
compatibly with \cite{Arkani-Hamed:2019ymq}, where we inverted \eqref{FA4.1} and \eqref{FA4.2} to get $a\cdot q=i\frac{\epsilon(p_1,p_2,q,a)}{m_1m_2\sigma v}$.

It is instructive to show how to get the 1PM amplitude also by summing over the helicities of intermediate polarizations,
\begin{align}
\label{FA7}
i\mathcal{A}^{\mathrm{1PM}}(a;p_i,q_i)=-\frac{i}{q^2}\sum_{h=\pm}\mathcal{A}_3^{\mu}(a_1;p_1,q)\epsilon^h_{\mu}(q)\epsilon^{-h}_{\nu}(-q)\mathcal{A}_3^{\nu}(a_2;p_2,-q).
\end{align}
Using now the definition of $\epsilon'^{\mu}$, the previous equation becomes,
\begin{align}
\label{FA8}
i\mathcal{A}^{\mathrm{1PM}}(a;p_i,q_i)=-\frac{i}{q^2}\sum_{h=\pm}\mathcal{A}_3^{\mu}(p_1,q)\epsilon'^{h}_{\mu}(a_1,q)\epsilon'^{-h}_{\nu}(a_2,-q)\mathcal{A}_3^{\nu}(p_2,-q).
\end{align}
We introduce projectors onto positive and negative helicity states,
\begin{align}
\label{FA9}
&\Pi^{+}_{\mu\nu}(q)=\epsilon_{\mu}^{+}(q)\epsilon_{\nu}^{-}(-q)=\frac{1}{2}\eta_{\mu\nu}+\frac{1}{2}\frac{q_{\mu}k_{\nu}+q_{\nu}k_{\mu}}{q\cdot k}+\frac{i}{2}\frac{\epsilon_{\mu\nu}(q,k)}{q\cdot k},\\
&\Pi^{-}_{\mu\nu}(q)=\epsilon_{\mu}^{-}(q)\epsilon_{\nu}^{+}(-q)=\frac{1}{2}\eta_{\mu\nu}+\frac{1}{2}\frac{q_{\mu}k_{\nu}+q_{\nu}k_{\mu}}{q\cdot k}-\frac{i}{2}\frac{\epsilon_{\mu\nu}(q,k)}{q\cdot k},
\end{align}
where $k$ is an arbitrary reference null vector. They satisfy,
\begin{align}
\label{FA10}
\Pi^{\pm}_{\mu}{}^{\rho}(q)\Pi_{\rho}^{\pm}{}^{\nu}(q)=\Pi_{\mu}^{\pm}{}^{\nu}(q),\qquad\Pi^{\pm}_{\mu}{}^{\rho}(q)\Pi_{\rho}^{\mp}{}^{\nu}(q)=0,
\end{align}
\begin{align}
\label{FA11}
\Pi^{+}_{\mu\nu}(q)+\Pi^{-}_{\mu\nu}(q)=\eta_{\mu\nu}+\frac{q_{\mu}k_{\nu}+q_{\nu}k_{\mu}}{q\cdot k}=\Pi_{\mu\nu}(q),\qquad \Pi_{\mu}{}^{\rho}(q)\Pi_{\rho}{}^{\nu}(q)=\Pi_{\mu}{}^{\nu}(q).
\end{align}
Up to gauge transformations $\epsilon'^{\mu}_h=e^{-ha\cdot q}\epsilon^{\mu}_h$ and hence the sum appearing in \eqref{FA8} reads,
\begin{align}
\label{FA12}
\nonumber \sum_{h=\pm}e^{-ha_1\cdot q}\epsilon^h_{\mu}(q)e^{ha_2\cdot q}\epsilon^h_{\nu}(-q)=&\bigg(\eta_{\mu\nu}-\frac{q_{\mu}k_{\nu}+q_{\nu}k_{\mu}}{q\cdot k}\bigg)\cosh(a\cdot q)\\&-i\frac{\epsilon_{\mu\nu}(q,k)}{q\cdot k}\sinh(a\cdot q)\equiv \Pi_{\mu\nu}(a;q),
\end{align}
where $a=a_1+a_2$. $\Pi_{\mu\nu}(a)$ is a spin-dressed projector and it satisfies,
\begin{align}
\label{FA13}
\Pi_{\mu}{}^{\rho}(a;q)\Pi_{\rho}{}^{\nu}(b;q)=\Pi_{\mu}{}^{\nu}(a+b;q).
\end{align}
The 1PM amplitude in \eqref{FA7} can be written in terms of $\Pi_{\mu}{}^{\nu}(a;q)$ as,
\begin{align}
\label{FA14}
i\mathcal{A}^{1\mathrm{PM}}(a;p_i,q_i)=-\frac{i}{q^2}\mathcal{A}_3^{\mu}(p_1,q)\Pi_{\mu\nu}(a;q)\mathcal{A}_3^{\nu}(p_2,-q).
\end{align}
This equation shows that the full dependence on the classical spin of the 1PM amplitude is actually encoded into a suitable ``spin dressing" of the Feynman propagator. 

Similarly, the gravity 1PM amplitude in \eqref{F61} can be written as,
\begin{align}
\label{FA15}
\nonumber i\mathcal{M}^{1\mathrm{PM}}(a;p_i,q_i)&=-\frac{i}{q^2}\mathcal{M}_3^{\mu\nu}(a_1;p_1,q)\Pi_{\mu\nu;\rho\sigma}(q)\mathcal{M}_3^{\rho\sigma}(a_2;p_2,-q)\\&=-\frac{i}{q^2}\mathcal{M}_3^{\mu\nu}(p_1,q)\Pi_{\mu\nu;\rho\sigma}(a;q)\mathcal{M}_3^{\rho\sigma}(p_2,-q),
\end{align}
where $\Pi_{\mu\nu;\rho\sigma}(a;q)$ is a spin-dressed de Donder projector,
\begin{align}
\label{FA16}
\nonumber \Pi_{\mu\nu;\rho\sigma}(a;q)=&\frac{1}{2}\bigg[\Pi_{\mu\rho}(a;q)\Pi_{\nu\sigma}(a;q)+\Pi_{\mu\sigma}(a;q)\Pi_{\nu\rho}(a;q)-\Pi_{\mu\nu}(a;q)\Pi_{\rho\sigma}(a;q)\bigg]\\=&\frac{1}{2}\bigg[\Pi_{\mu\rho}(q)\Pi_{\nu\sigma}(q)+\Pi_{\mu\sigma}(q)\Pi_{\nu\rho}(q)-\Pi_{\mu\nu}(q)\Pi_{\rho\sigma}(q)\bigg]\cosh(a\cdot q)\\\nonumber&-\frac{i}{2q\cdot k}\bigg[\epsilon_{\nu\sigma}(q,k)\Pi_{\mu\rho}(q)+\epsilon_{\mu\rho}(q,k)\Pi_{\nu\sigma}(q)\bigg]\sinh(a\cdot q).
\end{align}
satisfying,
\begin{align}
\label{FA17}
\Pi_{\mu\nu}{}^{\alpha\beta}(a;q)\Pi_{\alpha\beta}{}^{\rho\sigma}(b;q)=\Pi_{\mu\nu}{}^{\rho\sigma}(a+b;q).
\end{align}
\section{One-loop integrals}
\label{1-loop}
For the 2PM amplitude we need to solve the integral $\mathcal{I}^{\mu_1\dots\mu_n}_{\triangleright}(a)$ in \eqref{F78} and hence we start by considering the family of integrals,
\begin{align}
\mathcal{I}^{\mu_1\dots\mu_n}_{\triangleright}(0)=\int\frac{d^4l}{(2\pi)^4}\frac{l^{\mu_1}\dots l^{\mu_n}}{l^2(l+q)^2[(p_1+l)^2-m_1^2]}.
\end{align}
We find, 
\begin{align}
\label{FB1}
&\mathcal{I}_{\triangleright}(0)=-\frac{i}{32m_1\sqrt{-q^2}}+\dots,
\end{align}
and using the Passarino-Veltmano reduction \cite{Campbell:1996zw,Denner:2005nn,Fleischer:2010sq,Bjerrum-Bohr:2002gqz},
\begin{align}
\label{FB2}
&\mathcal{I}^{\mu}_{\triangleright}(0)=-\frac{i}{32m_1\sqrt{-q^2}}\bigg[-\frac{1}{2}q^{\mu}+\frac{1}{4}\frac{q^2}{m_1^2}p_1^{\mu}+\dots\bigg],\\
\label{FB3}
&\mathcal{I}^{\mu\nu}_{\triangleright}(0)=-\frac{i}{32m_1\sqrt{-q^2}}\bigg[\frac{3}{8}q^{\mu}q^{\nu}+\frac{1}{8}\frac{q^2}{m_1^2}p^{\mu}_1p_1^{\nu}-\frac{3}{8}\frac{q^2}{m_1^2}p_1^{(\mu}q^{\nu)}-\frac{1}{8}q^2\eta^{\mu\nu}+\dots\bigg],\\
\label{FB4}
\nonumber &\mathcal{I}^{\mu\nu\rho}_{\triangleright}(0)=-\frac{i}{32m_1\sqrt{-q^2}}\bigg[-\frac{5}{16}q^{\mu}q^{\nu}q^{\rho}+\frac{3}{16}\frac{q^4}{m_1^4}p^{\mu}_1p_1^{\nu}p_1^{\rho}-\frac{3}{16}\frac{q^2}{m_1^2}q^{(\mu}p^{\rho}_1p^{\nu)}_1\\&\hspace{3cm}+\frac{15}{32}\frac{q^2}{m_1^2}q^{(\mu}q^{\nu}p_1^{\rho)}+\frac{3}{16}q^2\eta^{(\mu\nu}q^{\rho)}-\frac{3}{16}\frac{q^2}{m_1^2}\eta^{(\mu\nu}p_1^{\rho)}+\dots\bigg],
\end{align}
where $\dots$ denote subleading terms in the classical limit. Expanding in powers of the total spin vector $a^{\mu}$ the integral $\mathcal{I}^{\mu_1\dots\mu_n}_{\triangleright}(a)$, we see that we need to compute the contractions of the above tensors with $a_{\mu_1}\dots a_{\mu_n}$. With the aligned spin and the kinematics specified in \eqref{F79}, all the scalar products $a\cdot p_i$ vanish. Hence, we are only interested in keeping the terms in \eqref{FB2}-\eqref{FB4} having products of $q$'s and $\eta$'s. For the latter there is a general formula \cite{Smirnov:2002pj},
\begin{align}
\label{FB5}
\nonumber \mathcal{I}^{\mu_1\dots\mu_n}(0)=&-\frac{i}{32m_1\pi^{\frac{3}{2}}\sqrt{-q^2}}\sum_{m=0}^{\left[\frac{n}{2}\right]}\frac{(-1)^n\Gamma(\frac{1}{2}-m)\Gamma(n-m+\frac{1}{2})\Gamma(m+\frac{1}{2})}{\Gamma(n+1)}\left(\frac{q^2}{2}\right)^{m}\\&\{[\eta]^m[q]^{n-2m}\}^{\mu_1\dots\mu_n},
\end{align}
in the classical limit, where the symbol $\{[\eta]^m[q]^{n-2m}\}^{\mu_1\dots\mu_n}$ comprises all the symmetric structures one can construct with $q$ and $\eta$ having $n$ covariant indices, \textit{e.g.}
\begin{align*}
&\{[\eta]^0[q]^{1}\}^{\mu_1}=q^{\mu_1},\\
&\{[\eta]^1[q]^{0}\}^{\mu_1\mu_2}=\eta^{\mu_1\mu_2},\\
&\{[\eta]^1[q]^{1}\}^{\mu_1\mu_2\mu_3}=\eta^{\mu_1\mu_2}q^{\mu_3}+\eta^{\mu_3\mu_1}q^{\mu_2}+\eta^{\mu_2\mu_3}q^{\mu_1}.
\end{align*}
We have, expanding $\cosh(x)$,
\begin{align}
\label{FB6}
\nonumber\mathcal{I}_{\triangleright}(a)=&\sum_{n=0}^{\infty}\frac{1}{(2n)!}\int\frac{d^4l}{2\pi)^4}\frac{(2a\cdot l+a\cdot q)^{2n}}{l^2(l+q)^2[(p_1+l)^2-m_1^2]}\\\nonumber =&\sum_{n=0}^{\infty}\sum_{k=0}^{2n}\frac{(a\cdot q)^{k}2^{2n-k}}{(2n-k)!k!}a_{\mu_1}\dots a_{\mu_{2n-k}}\int\frac{d^4l}{(2\pi)^4}\frac{l^{\mu_1}\dots l^{\mu_{2n-k}}}{l^2(l+q)^2[(p_1+l)^2-m_1^2]}\\\nonumber=&-\frac{i}{32m_1\pi^{\frac{3}{2}}\sqrt{-q^2}}\sum_{n=0}^{\infty}\sum_{k=0}^{2n}\sum_{m=0}^{[n-\frac{k}{2}]}\frac{(-1)^k2^{2n-k}}{(2n-k)!k!}\frac{\Gamma(\frac{1}{2}-m)\Gamma(2n-k+\frac{1}{2})\Gamma(m+\frac{1}{2})}{\Gamma(2n-k+1)}\\&\left(\frac{q^2 a^2}{2}\right)^m(a\cdot q)^{2n-2m}\frac{(2n-k)!}{m!(2n-k-2m)!2^m},
\end{align}
where we used that,
\begin{align}
\label{FB7}
a_{\mu_1}\dots a_{\mu_n}\{[\eta]^m[q]^{n-2m}\}^{\mu_1\dots\mu_n}=a^{2m}(a\cdot q)^{n-2m}\frac{n!}{m!(n-2m)!2^m}.
\end{align}
For $n=0,1,2$ the sum in \eqref{FB6} gives,
\begin{align}
\label{FB8}
-\frac{i}{32m_1\sqrt{-q^2}},\qquad-\frac{i[(a\cdot q)^2-a^2q^2]}{128m_1\sqrt{-q^2}},\qquad -\frac{i[(a\cdot q)^2-a^2
q^2]^2}{2048m_1\sqrt{-q^2}}.
\end{align}
These quantities are related to the series expansion of the modified Bessel function of the first kind $I_0(\sqrt{(a\cdot q)^2-a^2q^2})$ and, in particular we have,
\begin{align}
\label{FB9}
\mathcal{I}_{\triangleright}(a)=\int\frac{d^4l}{(2\pi)^4}\frac{\cosh(2a\cdot l+a\cdot q)}{l^2(l+q)^2[(p_1+l)^2-m_1^2]}=-\frac{i}{32 m_1\sqrt{-q^2}}I_0(\mathscr{F}(a,q)),
\end{align}
where we introduced for convenience,
\begin{align}
\label{FB9.1}
\mathscr{F}(a,q)=\sqrt{(a\cdot q)^2-a^2q^2},
\end{align}
in the aligned spin case, which is shift-symmetric \cite{Aoude:2022trd}. We could also derive,
\begin{align}
\label{FB9.15}
&a_{\mu}\mathcal{I}^{\mu}_{\triangleright}(a)=\frac{i}{32m_1\sqrt{-q^2}}\frac{(a\cdot q)}{2}I_0(\mathscr{F}(a,q)),\\\nonumber
&a_{\mu}a_{\nu}\mathcal{I}^{\mu\nu}_{\triangleright}(a)=-\frac{i}{32m_1\sqrt{-q^2}}\frac{1}{4}\bigg\{[\mathscr{F}^2(a,q)+(a\cdot q)^2]I_0(\mathscr{F}(a,q))\\&\hspace{2.35cm}-\mathscr{F}(a,q)I_1(\mathscr{F}(a,q))\bigg\}.
\end{align}
Using the above results, it is simple to show that the Passarino-Veltman decomposition generalizes also to the case of integrals of the form \eqref{F78} and, in particular, we find,
\begin{align}
\label{FB9.2}
&\mathcal{I}_{\triangleright}^{\mu}(a)=-\frac{i}{32m_1\sqrt{-q^2}}\bigg(-\frac{1}{2}q^{\mu}+\frac{1}{4}\frac{q^2}{m_1^2}p_1^{\mu}-\frac{1}{8}\frac{q^4}{m_1^2}\frac{a\cdot q}{\mathscr{F}(a,q)^2}a^{\mu}\bigg)I_0(\mathscr{F}(a,q)),\\
\label{FB9.3}
&\mathcal{I}_{\triangleright}^{\mu\nu}(a)=Aq^{\mu}q^{\nu}+Bq^{(\mu}p_1^{\nu)}+Cp_1^{\mu}p_1^{\nu}+D\eta^{\mu\nu}+Ea^{\mu}a^{\nu}+Fa^{(\mu}q^{\nu)}+Ga^{(\mu}p_1^{\nu)},
\end{align}
where,
\begin{align}
\label{FB9.25}
A=&-\frac{i}{128m_1\sqrt{-q^2}}\bigg[\frac{2(a\cdot q)^2-a^2q^2}{\mathscr{F}^2(a,q)}I_0(\mathscr{F}(a,q))-\frac{(a\cdot q)^2+a^2q^2}{\mathscr{F}(a,q)^3}I_1(\mathscr{F}(a,q))\bigg],\\
 B=&\frac{i}{128m_1\sqrt{-q^2}}\frac{q^2}{m_1^2}\bigg[\frac{2(a\cdot q)^2-a^2q^2}{\mathscr{F}(a,q)^2}I_0(\mathscr{F}(a,q))-\frac{(a\cdot q)^2+a^2q^2}{\mathscr{F}(a,q)^3}I_1(\mathscr{F}(a,q)\bigg],\\ C=&-\frac{i}{128m_1\sqrt{-q^2}}\frac{q^2}{m_1^2}\bigg[\frac{q^2}{2m_1^2}\frac{2(a\cdot q)^2-a^2q^2}{\mathscr{F}(a,q)^2}I_0(\mathscr{F}(a,q))+\frac{I_1(\mathscr{F}(a,q))}{\mathscr{F}(a,q)}\bigg],\\ D=&\frac{i}{128m_1\sqrt{-q^2}}q^2\bigg[\frac{q^2}{4m_1^2}\frac{2(a\cdot q)^2-a^2q^2}{\mathscr{F}(a,q)^2}I_0(\mathscr{F}(a,q))+\frac{I_1(\mathscr{F}(a,q))}{\mathscr{F}(a,q)}\bigg],\\ E=&-\frac{i}{64m_1\sqrt{-q^2}}q^4\bigg[\frac{I_0(\mathscr{F}(a,q))}{2\mathscr{F}(a,q)^2}-\frac{I_1(\mathscr{F}(a,q))}{\mathscr{F}(a,q)^3}\bigg],\\ F=&\frac{i}{32m_1\sqrt{-q^2}}q^2(a\cdot q)\bigg[\frac{I_0(\mathscr{F}(a,q))}{2\mathscr{F}(a,q)^2}-\frac{I_1(\mathscr{F}(a,q))}{\mathscr{F}(a,q)^3}\bigg],\\ G=&-\frac{i}{64m_1\sqrt{-q^2}}\frac{q^4(a\cdot q)}{m_1^2}\bigg[\frac{I_0(\mathscr{F}(a,q))}{2\mathscr{F}(a,q)^2}-\frac{I_1(\mathscr{F}(a,q))}{\mathscr{F}(a,q)^3}\bigg].
\end{align}
Using similar arguments we find,
\begin{align}
\label{FB10}
\mathcal{J}_{\triangleright}(a)=\int\frac{d^4l}{(2\pi)^4}\frac{\sinh(2a\cdot l+a\cdot q)}{l^2(l+q)^2[(p_1+l)^2-m_1^2]}=0,
\end{align}
and,
\begin{align}
\label{FB10.05}
&a_{\mu}\mathcal{J}^{\mu}_{\triangleright}(a)=-\frac{i}{64m_1\sqrt{-q^2}}\mathscr{F}(a,q)I_1(\mathscr{F}(a,q)),\\
&a_{\mu}a_{\nu}\mathcal{J}_{\triangleright}^{\mu}(a)=\frac{i}{64m_1\sqrt{-q^2}}(a\cdot q)\mathscr{F}(a,q)I_1(\mathscr{F}(a,q)).
\end{align}
The Passarino-Veltman decomposition now is,
\begin{align}
\label{FB10.1}
&\mathcal{J}^{\mu}_{\triangleright}(a)=-\frac{i}{64m_1\sqrt{-q^2}}\bigg[(a\cdot q)q^{\mu}-\frac{q^2(a\cdot q)}{2m_1^2}p_1^{\mu}-\frac{q^2}{2}a^{\mu}\bigg]\frac{I_1(\mathscr{F}(a,q))}{\mathscr{F}(a,q)},\\
\label{FB10.15}
&\mathcal{J}^{\mu\nu}_{\triangleright}(a)=Aq^{\mu}q^{\nu}+Bq^{(\mu}p_1^{\nu)}+Cp_1^{\mu}p_1^{\nu}+D\eta^{\mu\nu}+Ea^{\mu}a^{\nu}+Fa^{(\mu}q^{\nu)}+Ga^{(\mu}p_1^{\nu)},
\end{align}
with,
\begin{align}
\label{FB10.2}
&A=\frac{i(a\cdot q)}{64 m_1\sqrt{-q^2}}\frac{I_1(\mathscr{F}(a,q))}{\mathscr{F}(a,q)},\\
&B=-\frac{i(a\cdot q)}{64m_1\sqrt{-q^2}}\frac{q^2}{m_1^2}\frac{I_1(\mathscr{F}(a,q))}{\mathscr{F}(a,q)},\\
&C=\frac{i(a\cdot q)}{64m_1\sqrt{-q^2}}\frac{2q^4}{m_1^4}\frac{I_1(\mathscr{F}(a,q))}{\mathscr{F}(a,q)},\\
&D=-\frac{i(a\cdot q)}{64m_1\sqrt{-q^2}}\frac{q^4}{4m_1^2}\frac{I_1(\mathscr{F}(a,q))}{\mathscr{F}(a,q)},\\
&E=-\frac{i(a\cdot q)}{64m_1\sqrt{-q^2}}\frac{q^6}{2m_1^2}\frac{I_1(\mathscr{F}(a,q))}{\mathscr{F}(a,q)^3},\\
&F=-\frac{i}{64m_1\sqrt{-q^2}}q^2\frac{I_1(\mathscr{F}(a,q))}{\mathscr{F}(a,q)},\\
&G=\frac{i}{64m_1\sqrt{-q^2}}\frac{q^4}{2m_1^2}\frac{I_1(\mathscr{F}(a,q))}{\mathscr{F}(a,q)}.
\end{align}
We stress again here that the results for the various integrals above are valid only if $a\cdot p_1=0$. Furthermore, it is straightforward to prove the following relations,
\begin{align}
\label{FB11}
&\frac{\partial\mathcal{I}_{\triangleright}(a)}{\partial a_{\mu}}=2\mathcal{J}^{\mu}_{\triangleright}(a)+q^{\mu}\mathcal{J}_{\triangleright}(a),\\
\label{FB11.a}
&\frac{\partial^2 \mathcal{I}_{\triangleright}(a)}{\partial a_{\mu}\partial a_{\nu}}=4\mathcal{I}_{\triangleright}^{\mu\nu}(a)+2q^{\mu}\mathcal{I}_{\triangleright}^{\nu}(a)+2q^{\nu}\mathcal{I}^{\mu}_{\triangleright}(a)+q^{\mu}q^{\nu}\mathcal{I}_{\triangleright}(a),\\\label{FB11.b}&\frac{\partial\mathcal{J}_{\triangleright}(a)}{\partial a_{\mu}}=2\mathcal{I}^{\mu}_{\triangleright}(a)+q^{\mu}\mathcal{I}_{\triangleright}(a),\\ 
&\label{FB11.c}\frac{\partial^2 \mathcal{J}_{\triangleright}(a)}{\partial a_{\mu}\partial a_{\nu}}=4\mathcal{J}_{\triangleright}^{\mu\nu}(a)+2q^{\mu}\mathcal{J}_{\triangleright}^{\nu}(a)+2q^{\nu}\mathcal{J}^{\mu}_{\triangleright}(a)+q^{\mu}q^{\nu}\mathcal{J}_{\triangleright}(a),
\end{align}
that are compatible with what we have found so far.
\section{2PM eikonal phase integrals}
\label{AC}
In order to compute the 2PM eikonal, we need to evaluate the integral,
\begin{align}
\label{FC1}
\mathcal{I}_n[\mathbf{a};\alpha]\equiv \int\frac{d^2\mathbf{q}}{(2\pi)^2}\frac{(\mathbf{a}\cdot \mathbf{q})^n}{(\mathbf{q}^2)^{\alpha}}e^{-i\mathbf{q}\cdot \mathbf{b}}=i^na^{i_1}\dots a^{i_n}\frac{\Gamma(1-\alpha)}{ 2^{2\alpha}\pi\Gamma(\alpha)}\frac{\partial^n}{\partial b^{i_1}\dots\partial b^{i_n}}(\mathbf{b}^2)^{\alpha-1}.
\end{align}
Using,
\begin{align}
\label{FC2}
\frac{\partial^n}{\partial b^{i_1}\dots\partial b^{i_n}}(\mathbf{b}^2)^{\alpha-1}=\sum_{m=0}^{[\frac{n}{2}]}\frac{2^{n-m}\Gamma(\alpha)}{\Gamma(\alpha-n+m)}(\mathbf{b}^2)^{\alpha-1-n+m}\{[\delta]^m[\mathbf{b}]^{n-2m}\}^{i_1\dots i_n},
\end{align}
and \eqref{FB7} we obtain,
\begin{align}
\label{FC3}
\mathcal{I}_n[\mathbf{a};\alpha]=\frac{1}{ 2^{2\alpha}\pi}\sum_{m=0}^{[\frac{n}{2}]}\frac{\Gamma(1-\alpha)(2i)^{n}n!}{2^{2m}\Gamma(\alpha-n+m)m!(n-2m)!}(\mathbf{b}^2)^{\alpha-1-n+m}(\mathbf{a}^2)^m(\mathbf{a}\cdot \mathbf{b})^{n-2m}.
\end{align}
We are interested in computing the Fourier transform of modified Bessel functions appearing in the 2PM amplitude \eqref{F80} and \eqref{F86}. We have,
\begin{align}
\label{FC4}
\nonumber\tilde{I}_0[\mathbf{a},\mathbf{b}]&=\int\frac{d^2\mathbf{q}}{(2\pi)^2}\frac{e^{-i\mathbf{b}\cdot\mathbf{q}}}{(\mathbf{q}^2)^{\frac{1}{2}}}I_0(\mathscr{F}(\mathbf{a},\mathbf{q}))=\sum_{n=0}^{\infty}\frac{1}{2^{2n}(n!)^2}\int\frac{d^2\mathbf{q}}{(2\pi)^2}\frac{e^{-i\mathbf{b}\cdot\mathbf{q}}}{(\mathbf{q}^2)^{\frac{1}{2}}}[(\mathbf{a}\cdot\mathbf{q})^2-\mathbf{a}^2\mathbf{q}^2]^n\\\nonumber &=\sum_{n=0}^{\infty}\sum_{k=0}^n\frac{(-1)^k(\mathbf{a}^2)^{k}}{2^{2n}n!k!(n-k)!}a_{i_1}\dots a_{i_{2n-2k}}\int\frac{d^2\mathbf{q}}{(2\pi)^2}\frac{e^{-i\mathbf{b}\cdot\mathbf{q}}}{(\mathbf{q}^2)^{\frac{1}{2}-k}}q^{i_1}\dots q^{i_{2n-2k}}\\&=\sum_{n=0}^{\infty}\sum_{k=0}^n\frac{(-1)^k(\mathbf{a}^2)^{k}}{2^{2n}n!k!(n-k)!}\mathcal{I}_{2n-2k}[\mathbf{a};1/2-k].
\end{align}
Similarly,
\begin{align}
\label{FC5}
\nonumber\tilde{\bigg(\frac{I_1}{\mathscr{F}}\bigg)}[\mathbf{a},\mathbf{b}]&=\int\frac{d^2\mathbf{q}}{(2\pi)^2}\frac{e^{-i\mathbf{q}\cdot\mathbf{b}}}{(\mathbf{q}^2)^\frac{1}{2}}\frac{I_1(\mathscr{F}(\mathbf{a},\mathbf{q}))}{\mathscr{F}(\mathbf{a},\mathbf{q})}\\&=\sum_{n=0}^{\infty}\sum_{k=0}^n\frac{(-1)^k(\mathbf{a}^2)^{k}}{2^{2n+1}(n+1)!k!(n-k)!}\mathcal{I}_{2n-2k}[\mathbf{a};1/2-k],
\end{align}
and,
\begin{align}
\label{FC6}
\nonumber\tilde{I}_1[\mathbf{a},\mathbf{b}]&=\int\frac{d^2\mathbf{q}}{(2\pi)^2}\frac{e^{-i\mathbf{b}\cdot\mathbf{q}}}{(\mathbf{q}^2)^{\frac{1}{2}}}I_1(\mathscr{F}(\mathbf{a},\mathbf{q}))\\&=\sum_{n,l=0}^{\infty}\sum_{k=0}^n\frac{(-1)^k(2l)!(\mathbf{a}^2)^{l+k}}{2^{2n+2l+1}(n+1)!(1-2l)(l!)^2k!(n-k)!}\mathcal{I}_{2n+1-2k-2l}[\mathbf{a};1/2-k-l].
\end{align}
\bibliographystyle{utphys}
\bibliography{hie4.bib}

\end{document}